\documentclass{llncs}

\newif\iffinal
  \finaltrue

\newif\ifwithextensions
\withextensionsfalse

\usepackage{amsthm}
\usepackage{amsfonts}
\usepackage{amsmath}
\usepackage{amssymb}
\usepackage{temporal}
\usepackage{cite}
\usepackage{graphicx}
\usepackage{enumitem}
\usepackage{wrapfig}
\usepackage{environ}
\usepackage{multirow}
\usepackage{xspace}
\usepackage{color,soul}
  \definecolor{lightblue}{rgb}{.8,.95,1}

\usepackage{thmtools, thm-restate}

\usepackage{tikz}
\usepackage{pgf}
\usetikzlibrary{positioning,calc,automata,arrows,fit,shapes}

\usepackage{caption}
\usepackage{mdframed}

\let\note\relax

\iffinal
  \usepackage[finalold]{trackchanges} 
\else
  \usepackage[inline]{trackchanges}
\fi

\addeditor{$S\!J$}
\addeditor{$M\!S$}
\addeditor{$A\!K$}
\newcommand{\sj}[1]{\note[$S\!J$]{#1}}
\newcommand{\ms}[1]{\note[$M\!S$]{#1}}

\DeclareTextFontCommand{\emph}{\em}

\newcommand{\listOfKeywords}{parameterized model checking, reactive synthesis, distributed systems, guarded protocols, deadlocks, fairness}

\newcommand{\mailto}[1]{\href{mailto:#1}{\nolinkurl{#1}}}

\newcommand{\mC}{\mathcal{C}}

\newcommand{\mN}{\mathcal{N}}

\newcommand{\mP}{\mathcal{P}}
\newcommand{\mB}{\mathcal{B}}

\newcommand{\init}{{\sf init}\xspace}
\newcommand{\templates}{P}

\newcommand{\inputs}{\Sigma}
\newcommand{\localin}{\sigma}
\newcommand{\globIn}{E}

\newcommand{\visited}{\mathsf{Visited}}

\newcommand{\first}{\mathsf{first}}

\newcommand{\sched}{{{\sf move}}}
\newcommand{\enabled}{{{\sf en}}}

\newcommand\LTLmX{\ensuremath{\mbox{\textsf{LTL}}\backslash\textsf{X}}\xspace}
\newcommand\LTL{\ensuremath{\mbox{\textsf{LTL}}}\xspace}

\definecolor{darkgreen}{rgb}{0,0.5,0}
\definecolor{darkblue}{rgb}{0,0,.5}
\definecolor{mygray}{gray}{.3}

\newcommand{\state}{q}
\newcommand{\initstate}{{\sf init}\xspace}
\newcommand{\stateset}{\expandafter\MakeUppercase\expandafter{\state}}
\newcommand{\State}{s}
\newcommand{\Stateset}{\expandafter\MakeUppercase\expandafter{\State}}

\newcommand{\trans}{\ensuremath{\delta}}
\newcommand{\Trans}{\ensuremath{\Delta}}

\renewcommand{\time}{m}
\newcommand{\dead}{{\sf dead}\xspace}
\newcommand{\Enable}{{\sf Enable}\xspace}
\newcommand{\appears}{{\sf appears}\xspace}
\newcommand{\last}{{\sf last}\xspace}
\newcommand{\occurs}{{\sf occurs}\xspace}

\newcommand{\card}[1]{\left| {#1} \right|}
\newcommand{\Nat}{\ensuremath{\mathbb{N}}}

\newcommand{\cupdot}{\mathbin{\dot{\cup}}}

\newcommand{\smartpar}[1]{\medskip \noindent {\bf #1}}

\newcommand{\spec}{\Phi}

\newcommand{\impl}{\rightarrow}

\declaretheorem[name=Observation]{obs}

\newcommand{\cutoffsys}{\ensuremath{(A,B)^{(1,c)}}\xspace}
\newcommand{\largesys}{\ensuremath{(A,B)^{(1,n)}}\xspace}

\iffinal
\newcommand{\gray}[1]{}

\else
\newcommand{\gray}[1]{{\color{black!50} #1}}

\fi

\newcommand{\li}{\begin{itemize}}
\newcommand{\il}{\end{itemize}}

\NewEnviron{tikzLTS}{%
  \begin{tikzpicture}[node distance=2cm,inner sep=1pt,minimum size=0.5mm,bend angle=20]
	 	\tikzstyle{proc} = [rectangle,draw=black,fill=green!20,thick,inner sep=10pt]
	  	\tikzstyle{state} = [circle,draw=black,thick,inner sep=3pt]
	  	\tikzstyle{noproc} = [circle]
	  	\tikzstyle{lbl} = [rectangle,node distance=2cm]
  		\tikzstyle{pre} = [ <-,shorten <=2pt,shorten >=2pt, >=stealth', semithick]
	  	\tikzstyle{post} = [ ->,shorten <=2pt,shorten >=2pt, >=stealth', semithick]
    \BODY
  \end{tikzpicture}
}

\NewEnviron{tikzPath}{%
  \begin{tikzpicture}	
  	\tikzstyle{pstate} = [circle, fill=black,thick, inner sep=2pt, minimum size=2mm]
    \tikzstyle{hiddenstate} = [circle]
  	\tikzstyle{trans-notsched} = [ ->,shorten <=2pt,shorten >=2pt, >=stealth', semithick]
  	\tikzstyle{trans-sched} = [ ->,shorten <=2pt,shorten >=2pt, >=stealth', very thick] 
    \BODY
    \end{tikzpicture}
}

\usepackage{tikz}
\usetikzlibrary{arrows,shapes,decorations,automata,backgrounds,petri}
\usepackage[colorlinks=true]{hyperref}
\hypersetup{%
  pdftitle={Analyzing Guarded Protocols},%
  pdfauthor={Swen Jacobs, Mouhammad Sakr},%
  pdfsubject={},%
  pdfkeywords={\listOfKeywords},%
  colorlinks%
}

\usepackage{booktabs}

\title{Analyzing Guarded Protocols: Better Cutoffs, More Systems, More Expressivity}
\titlerunning{Analyzing Guarded Protocols: Better Cutoffs, More Systems, More Expressivity}
\author{Swen Jacobs\inst{1}, Mouhammad Sakr\inst{1}}
\institute{$^1$ Reactive Systems Group, Saarland University, Germany}
\authorrunning{S. Jacobs and M. Sakr}

\begin{document}
%
%
%
%

\maketitle              
 \begin{abstract} 
We study cutoff results for parameterized verification and synthesis of 
guarded protocols, as introduced by Emerson and Kahlon (2000). Guarded 
protocols describe systems of processes whose transitions are enabled or 
disabled depending on the existence of other processes in certain local 
states. Cutoff results reduce reasoning about systems with an arbitrary 
number of processes to systems of a determined, fixed size. Our work is based 
on the observation that existing cutoff results for guarded protocols are 
often impractical, since they scale linearly in the number of local states of 
processes in the system. We provide new cutoffs that scale not with the 
number of local states, but with the number of guards in the system, which is 
in many cases much smaller. Furthermore, we consider natural extensions of the classes of systems and specifications under consideration, and present results for problems that have not been known to admit cutoffs before.
\end{abstract}


\sj{throughout whole paper: clarify the role and effect of process $A$!}

\section{Introduction}
\label{sec:intro}

Concurrent systems are notoriously hard to get correct, and are therefore a promising application area for formal methods like model checking or synthesis. However, while such general-purpose formal methods can give strong correctness guarantees, they have two drawbacks: i) the state explosion 
problem prevents us from using them for systems with a large number 
of components, and ii) correctness properties are often expected to hold 
for an \emph{arbitrary} number of components, which cannot be guaranteed without an additional argument that extends a proof of correctness to systems of arbitrary size. Both problems can be solved by approaches for 
\emph{parameterized} model checking and synthesis, which give correctness 
guarantees for systems with any number of components without considering every 
possible system instance explicitly.

While parameterized model checking (PMC) is undecidable even if we restrict systems to uniform finite-state components~\cite{Suzuki88}, 
there exist a number of methods that decide the problem for specific classes 
of systems~\cite{German92,EsparzaFM99,Emerson00,Emerso03,EmersonK03,Clarke04c,AJKR14}, some of which have been collected in surveys of the literature recently~\cite{Esparza14,BloemETAL15}. Additionally, there are semi-decision 
procedures that are successful in many interesting 
cases~\cite{Kurshan95,Bouajjani00,Clarke08,KaiserKW10}.
In this paper, we consider the \emph{cutoff} approach to PMC, that can guarantee properties of 
systems of arbitrary size by considering only systems of up to a certain 
fixed size, thus providing a decision procedure for PMC if components are finite-state.

Guarded protocols, the systems under consideration, are composed of an arbitrary number of processes, 
each an instance of a finite-state process template.
Processes communicate by guarded updates, 
where guards are statements about other processes that are interpreted either 
conjunctively (``every other process satisfies the guard'') or disjunctively 
(``there exists a process that satisfies the guard''). Conjunctive guards can 
be used to model atomic sections or locks, while disjunctive guards can model 
pairwise rendezvous or token-passing. 

This class of systems has been studied by Emerson and 
Kahlon~\cite{Emerson00,EmersonK03}, and cutoffs that depend on the 
size of process templates are known for specifications of the form 
$\forall{\bar{p}}.\ \spec(\bar{p})$, 
where $\spec(\bar{p})$ is an $\LTLmX$ property over the local states of one or more processes 
$\bar{p}$. 
Au{\ss}erlechner et al.~\cite{AJK16} have extended and improved these results, but a number of open issues remain. We will explain some of them in the following.

\paragraph*{Motivating Example}
\sj{can this be shortened?}

\begin{wrapfigure}{r}{0.4\linewidth}
\vspace{-8pt}
\scalebox{0.95}{
\input{reader-writer-tikz}
}
\label{fig:reader-writer}
\vspace{-18pt}
\end{wrapfigure}
As an example, consider the reader-writer protocol on the right, modeling access to data shared between processes. A process can signal that it wants to read the data by entering state $tr$ (``try-read''). From $tr$, it can move to the reading state $r$. However, this transition is guarded by a statement $\neg w$, meaning that no other process should currently be in state $w$, i.e., writing the data. Similarly, a process that wants to enter $w$ has to go through $tw$, and the transition into $w$ is guarded by $\neg w \land \neg r$, i.e., no state should be either reading or writing.

The cutoff results by Emerson and Kahlon~\cite{Emerson00} allow us to check parameterized safety conditions such as 
$$\forall i \neq j. \always \left( \neg (w_i \land w_j) \land \neg (w_i \land r_j) \right),$$
where indices $i$ and $j$ refer to different processes in the system.
In particular, they provide a cutoff that is linear in the size of the process template for detecting the absence of global deadlocks, and (assuming that deadlocks are not possible) an efficient cutoff of $2$ for $1$-indexed \LTLmX formulas, which can be generalized to a cutoff of $k+1$ for $k$-indexed properties.

However, when considering a liveness property such as
$$\forall i. \always \left( (tr_i \rightarrow \eventually r_i) \land (tw_i \rightarrow \eventually w_i) \right),$$
then their cutoff results are not very useful, since they do not consider fairness assumptions on the scheduling of processes, and there obviously exists a run with unfair scheduling that violates the property.

Au{\ss}erlechner et al.~\cite{AJK16} have looked at this problem, and divided it into two aspects: i) cutoffs for the detection of local deadlocks under the assumption of strong fairness, and ii) cutoffs for \LTLmX properties under the assumption of unconditional fairness. Since strong fairness and absence of local deadlocks imply unconditional fairness, this enables the verification of liveness properties under the assumption of strong fairness.
For ii), the provided cutoff is the same as for the non-fair case. For i), they give a cutoff that is linear in the size of the process template, but only for a restricted class of process templates. 

A number of limitations of the existing results is highlighted by the example above. First, the existing cutoff results for local deadlock detection do not support the given process template. More specifically, they only support $1$-conjunctive systems, i.e., systems where each guard can only exclude a single state. 
In this paper, we consider generalizations of this restricted class of process templates, and provide cutoffs for a class that includes examples such as the given one. Furthermore, we show that the general problem is very hard.

Another drawback of the existing results is that they use only minimal knowledge about the process templates: the size of templates and the type of guards. As a result, many cutoffs are linear in the size of the process template. Intuitively, the \emph{communication} between processes should be more important for the cutoff than their internal state space. This can be seen in the example above: out of the $5$ states, only $2$ can be observed by the other processes, and can thus influence their behavior. In this paper, we investigate how cutoff results change when we also consider communication-related measures of the process templates, such as the number of different guards, or the number of states that appear in guards.

\paragraph*{Contributions}
We provide new cutoff results for guarded protocols: 
\begin{enumerate}
\item We show that by closer analysis of process templates, in particular the number and the form of transition guards, we can get smaller cutoffs in many cases. This circumvents the tightness results of Au{\ss}erlechner et al.~\cite{AJK16}, which state that no smaller cutoffs can exist for the class of all processes of a given size. 
\item For conjunctive systems, we additionally extend the class of process templates that are supported by cutoff results. In particular, we provide cutoff results for local deadlock detection in classes of templates that are not $1$-conjunctive. However, we do not solve the general problem, and instead show that a cutoff for arbitrary conjunctive systems would at least be quadratic in the size of the template.
\item For disjunctive systems, we additionally extend both the class of process templates and the class of specifications that are supported by cutoff results. In particular, we show that systems with finite conjunctions of disjunctive guards are also supported by many of the existing proof methods, or variations of them. Based on this observation, we obtain cutoff results for these systems. Furthermore, we give cutoffs that support checking the simultaneous reachability (and repeated reachability) of a target set by \emph{all} processes in a disjunctive system. 
\end{enumerate}
\section{Preliminaries}
\label{sec:prelim}

\subsection{System Model}
\label{sec:model}

We consider systems $A {\parallel} B^n$, usually written $\largesys$, 
consisting of one copy of a process template $A$ and $n$ copies of a process template $B$, 
in an interleaving parallel composition.\footnote{Process template $A$ may be a trivial process that does nothing if we want to just consider a system $B^n$, as in the example in Section~\ref{sec:intro}.}%
We distinguish objects that belong to different templates by indexing them with
the template. E.g., for process template $U \in \{A,B\}$, $Q_U$ is the set of
states of $U$. For this section, fix two disjoint finite sets $Q_A$, $Q_B$ as
sets of states of process templates $A$ and $B$, and a positive integer $n$.

\smartpar{Processes.} A \emph{process template} 
 is a transition system
  $U=(\stateset, \init, \inputs, \trans)$ with 
	\begin{itemize}
	\item $\stateset$ is a finite set of states including the
  initial state $\init$,
	\item $\inputs$ is a finite input alphabet,
	\item $\trans: \stateset \times \inputs \times \mP(Q_A \cupdot Q_B) \times \stateset$ is a guarded transition relation.
	\end{itemize}
A process template is \emph{closed} if $\inputs = \emptyset$, and otherwise \emph{open}.

For  $U \in \{A,B\}$, define the size $\card{U} = \card{\stateset_U}$. We write $G_U$ for the set of non-trivial guards that are used in $\trans_U$, i.e., guards different from $Q_A \cup Q_B$ and $\emptyset$. Then, let $G = G_A \cup G_B$.

A copy of template $U$ will be called a \emph{$U$-process}.
Different $B$-processes are distinguished by subscript, i.e., for $i \in [1..n]$, $B_i$ is the $i$th copy of $B$, and $\state_{B_i}$ is a state of $B_i$. A state of the $A$-process is denoted by $q_A$. 

For the rest of this subsection, fix templates $A$ and $B$. We assume that $\inputs_A \cap \inputs_B = \emptyset$. We will also write $p$ for a process in $\{ A, B_1, \ldots, B_n\}$, unless $p$ is specified explicitly.

\smartpar{Disjunctive and Conjunctive Systems.}
In a system $\largesys$, consider global state $s = (\state_A,\state_{B_1},\ldots,\state_{B_n})$ and global input $e=(\localin_A,\localin_{B_1},\ldots,\localin_{B_n})$.
We also write $s(p)$ for $q_p$, and $e(p)$ for $\sigma_p$.
A local transition $(\state_p,\localin_p,g,\state_p') \in \trans_U$ of $p$ is \emph{enabled for $s$ and $e$} if its \emph{guard} $g$ is satisfied for $p$ in $s$, written $(s,p) \models g$. 
Disjunctive and conjunctive systems are distinguished by the \emph{interpretation of guards}:
\begin{align*}
\text{In disjunctive systems: } & (s,p) \models g \text{~~~iff~~~} 
\exists p' \in \{A,B_1,\ldots,B_n\} \setminus \{p\}:\ \ \state_{p'} \in g. \\
\text{In conjunctive systems: } & (s,p) \models g \text{~~~iff~~~} 
\forall p' \in \{A,B_1,\ldots,B_n\} \setminus \{p\}:\ \ \state_{p'} \in g.
\end{align*}

Note that we check containment in the guard (disjunctively or conjunctively) 
only for local states of processes \emph{different from} $p$. A process is \emph{enabled} for $s$ and $e$ if at least one of its transitions is enabled for $s$ and $e$, otherwise it is \emph{disabled}.

Like Emerson and Kahlon~\cite{Emerson00}, 
we assume that in conjunctive systems $\init_A$ and $\init_B$ are contained in all guards,
i.e., they act as neutral states.
For conjunctive systems, we call a guard \emph{$n$-conjunctive} if it is of the form $(Q_A \cupdot Q_B) \setminus \{q_1,\ldots,q_n\}$ for some $q_1,\ldots,q_n \in Q_A\cupdot Q_B$. A state $q$ is \emph{$1$-conjunctive} if all non-trivial guards of transitions from $q$ are $1$-conjunctive. A conjunctive system is \emph{$1$-conjunctive} if every state is $1$-conjunctive. 

Then, \largesys is defined as the transition 
system $(S,\init_S,\globIn,\Trans)$ with 
\begin{itemize}
\item set of global states $S = (\stateset_A) \times (\stateset_B)^{n}$, 
\item global initial state $\init_S = (\initstate_A,\initstate_B,\ldots,\initstate_B)$, 
\item set of global inputs $\globIn = (\inputs_A) \times (\inputs_B)^{n}$,
\item and global transition relation $\Trans \subseteq S \times \globIn \times S$ with $(s,e,s') \in \Trans$ iff 
\begin{enumerate}[label=\roman*)] 
  \item $s=(\state_A,\state_{B_1},\ldots,\state_{B_n})$, 
  \item $e=(\localin_A, \localin_{B_1},\ldots,\localin_{B_n})$, and 
  \item $s'$ is obtained from $s$ by replacing one local state $\state_p$ with a new local state $\state_p'$, where $p$ is a $U$-process with local transition $(\state_{p},\localin_{p},g,\state_p') \in \trans_U$ and $(s,p) \models g$. 
\end{enumerate}
\end{itemize}
We say that a system $\largesys$ is \emph{of type} $(A,B)$. 
A system is \emph{closed} if all of its templates are closed.
We often denote the set $\{B_1,...,B_n\}$ as $\mB$.

\smartpar{Runs.} 
A \emph{configuration} of a system is a triple $(s,e,p)$, where $s \in S$, $e 
\in \globIn$, and $p$ is either a system process, or the special symbol $\bot$.
 A \emph{path} of a system is a configuration sequence 
$x = (s_1,e_1,p_1),(s_2,e_2,p_2),\ldots$ such that for all $\time < |x|$ there is a 
transition $(s_\time,e_\time,s_{\time+1}) \in \Trans$ based on a local 
transition of process $p_\time$. We say that process 
$p_\time$ \emph{moves} at \emph{moment} $\time$. 
Configuration $(s,e,\bot)$ appears
 iff all processes are disabled for $s$ and $e$.
Also, for every $p$ and $\time < |x|$: 
either $e_{\time+1}(p) = e_\time(p)$ or process $p$ moves at moment $\time$. 
That is, the environment keeps input to each process unchanged until 
the process can read it.\footnote{By only considering inputs that are actually processed, we 
approximate an 
action-based semantics. Paths that do not fulfill this requirement are not 
very interesting, since the environment can violate any interesting 
specification that involves input signals by manipulating them when the 
corresponding process is not allowed to move.} 

A system \emph{run} is a maximal path starting in the initial state. Runs are either infinite, or they end in a configuration $(s,e,\bot)$. We say that a run is \emph{initializing} if every 
process
that moves infinitely often also visits 
its $\initstate$ 
infinitely often.

Given a system path $x = (s_1,e_1,p_1),(s_2,e_2,p_2),\ldots$ and a process $p$, the \emph{local path} of $p$ in $x$ is the projection $x(p) = (s_1(p),e_1(p)),(s_2(p),e_2(p)),\ldots$ of $x$ onto local states and inputs of $p$. $x(p)$ is a \emph{local run} if $x$ is a run.
Similarly define the projection on two processes $p_1,p_2$ denoted by $x(p_1,p_2)$.


\smartpar{Deadlocks and Fairness.}
A run is \emph{globally deadlocked} if it is finite.
An infinite run is \emph{locally deadlocked} for process $p$ if there exists $\time$ such that $p$ is disabled for all $s_{\time'},e_{\time'}$ with $\time'\ge \time$. A run is \emph{deadlocked} if it is locally or globally deadlocked.
A system \emph{has a (local/global) deadlock} if it has a (locally/globally) deadlocked run. Note that absence of local deadlocks for all $p$ implies absence of global deadlocks, but not the other way around.

A run $(s_1,e_1,p_1), (s_2,e_2,p_2),...$ is \emph{unconditionally-fair} if every process moves infinitely often. 
A run is \emph{strong-fair} if it is infinite and for every process $p$, if $p$ is enabled infinitely often, then $p$ moves infinitely often.

\subsection{Specifications}
\label{sec:semantics}
Fix templates $(A,B)$. We consider formulas in $\LTLmX$, i.e., $\LTL$ without the next-time operator $\nextt$.
Let $h(A,B_{i_1},\ldots,B_{i_k})$ be an $\LTLmX$ formula over atomic propositions from $Q_A \cup \Sigma_A$ and indexed propositions from $(Q_B \cup \Sigma_B) \times \{i_1,\ldots,i_k\}$. For a system $\largesys$ with $n \geq k$ and $i_j \in [1..n]$, satisfaction of $\pforall h(A,B_{i_1},\ldots,B_{i_k})$ and $\pexists h(A,B_{i_1},\ldots,B_{i_k})$ is defined in the usual way (see e.g. \cite{PrinciplesMC}).

\smartpar{Parameterized Specifications.} 	
\label{sec:parameterized}
A \emph{parameterized specification} is a temporal logic formula
with indexed atomic propositions and quantification over indices. 
A \emph{$k$-indexed formula} is of the form $\forall{i_1,\ldots,i_k.} \pforall h(A,B_{i_1},\ldots,B_{i_k})$ or $\forall{i_1,\ldots,i_k.} \pexists h(A,B_{i_1},\ldots,B_{i_k})$. 
For given $n \geq k$, 
$$\largesys {\models} \forall{i_1,{\ldots},i_k.} \pforall h(A,B_{i_1},{\ldots},B_{i_k})$$
~iff~
$$\largesys {\models} \bigwedge_{j_1 \neq {\ldots} \neq j_k \in [1..n]} \pforall h(A,B_{j_1},{\ldots},B_{j_k}).$$ 
By symmetry of guarded protocols, this is equivalent 
(cp.\cite{Emerson00})
to $\largesys \models \pforall h(A,B_1,\ldots,B_k)$. 
The latter formula is denoted by $\pforall h(A,B^{(k)})$, 
and we often use it instead of the original $\forall{i_1,\ldots,i_k.} \pforall h(A,B_{i_1},...,B_{i_k})$. For formulas with path quantifier $\pexists$, satisfaction is defined analogously, and equivalent to satisfaction of $\pexists h(A,B^{(k)})$.

\smartpar{Specification of Fairness and Local Deadlocks.}
It is often convenient to express fairness assumptions and local deadlocks 
as parameterized specifications.
To this end,
define auxiliary atomic propositions $\sched_p$ and $\enabled_p$ for every process $p$ of system $(A,B)^{(1,n)}$. At moment $\time$ of a given run $(s_1,e_1,p_1),(s_2,e_2,p_2), \ldots$, let $\sched_p$ be true whenever $p_\time = p$, and let $\enabled_p$ be true if $p$ is enabled for $s_\time, e_\time$. Note that we only allow the use of these propositions to define fairness, but not in general specifications.
Then, an infinite run is 
\begin{itemize}
\item \emph{local-deadlock-free} if it satisfies $\forall{p}. \GF \enabled_p$, abbreviated as $\spec_{\neg dead}$,
\item \emph{strong-fair} if it satisfies $\forall{p}. \GF \enabled_p \impl \GF \sched_p$, abbreviated as $\spec_{strong}$, and 
\item \emph{unconditionally-fair} if it satisfies $\forall{p}. \GF \sched_p$, abbreviated as $\spec_{uncond}$.
\end{itemize}


\subsection{Model Checking Problems and Cutoffs}
\label{sec:nonparameterized_synthesis}
For a given system $\largesys$ and specification $h(A,B^{(k)})$ with $n \ge k$,
\begin{itemize}
\item the \emph{model checking problem} is to decide whether $\largesys \models \pforall h(A,B^{(k)})$,
\item the (global/local) \emph{deadlock detection problem} is to decide whether $\largesys$
      has (global/local) deadlocks,
\item the \emph{parameterized model checking problem} (PMCP) is to decide whether $\forall m \ge n:\ (A,B)^{(1,m)} \models \pforall h(A,B^{(k)})$, and 
\item the \emph{parameterized (local/global) deadlock detection problem} is to decide whether 
      for some $m \ge n$, $(A,B)^{(1,m)}$ does have (global/local) local deadlocks.
\end{itemize}

These definitions can be flavored with different notions of fairness, and with the $\pexists$ path quantifier instead of $\pforall$.
Also, corresponding problems for the \emph{synthesis} of process templates can be defined (compare Au{\ss}erlechner et al.~\cite{AJK16}). Parameterized synthesis based on cutoffs~\cite{Jacobs14} is also supported by our cutoff results, but the details will not be necessary for understanding the results presented here.


\paragraph{Cutoffs.}
We define cutoffs with respect to a class of systems (either disjunctive or conjunctive), a class of process templates $\templates$, and a class of properties, which can be $k$-indexed formulas for some $k \in \Nat$ or the existence of (local/global) deadlocks. 

A \emph{cutoff} for a given class of properties and a class of systems with processes from $\templates$ is a number $c \in \Nat$ such that for all $A,B \in \templates$ and all properties $\varphi$ in the given class: 
$$ \largesys \models \varphi ~\Leftrightarrow~ \cutoffsys \models \varphi.$$
%

Like the problem definitions above, cutoffs may additionally be flavoured with different notions of fairness.

\paragraph{Cutoffs and Decidability.} Note that the existence of a cutoff implies that the parameterized model checking and parameterized deadlock detection problems are \emph{decidable} iff their non-parameterized versions are decidable.
\section{Better Cutoffs for Disjunctive Systems}
In  this section, we state our new cutoff results for disjunctive systems, and compare them to the previously known results in Table~\ref{tab:cutoffs-disj}.
Full proofs can be found in Appendix~\ref{sec:proofs-disj}.

To state our first theorem, we need the following additional definitions.

Fix process templates $A,B$ with $G=G_A \cup G_B$. Let $|B|_G = |\{q \in Q_B \mid \exists g \in G: q \in g\}|$. For a state $q \in Q_B$ in a disjunctive system, define $\Enable_q = \{ q' \in Q_A \cup Q_B \mid \exists (q,\sigma,g,q'') \in \trans_B: q' \in g\}$, i.e., the set of states of $A$ and $B$ that enable a transition from $q$. Furthermore, let $\mN = \{q \in Q_B \mid q \in \Enable_q\}$, and let $\mN^*$ be the maximal subset (wrt. number of elements) of $\mN$ such that $\forall q_i,q_j \in \mN^*: q_i \notin \Enable_{q_j} \land q_j \notin \Enable_{q_i}$.
Then we obtain:

\begin{theorem}[Disjunctive Cutoff Theorem]
\label{thm:disj}
For disjunctive systems and process templates $A,B$ with $G = G_A \cup G_B$: \sj{can we improve more results by using $|B|_G$?}
\begin{itemize}
\item $|B|_G+k+1$ and $|G| + k + 1$ are cutoffs for $k$-indexed properties in non-fair executions,
\item $|B| + |G| + k$ is a cutoff for $k$-indexed properties in unconditionally fair executions,
\item $m + |G| + 1$ is a cutoff for local deadlock detection in non-fair executions, where $m = \max_{q \in Q^*_B}\{|\Enable_q|\}$ for $Q^*_B=\{q \in Q_B \mid |\Enable_q| < |B|\}$,
\item $|B| + |G|$ is a cutoff for local deadlock detection in unconditionally fair executions,
\item $|B| + |\mN^*|$ is a cutoff for global deadlock detection. \sj{can deadlock cutoffs be improved by removing states that are both free and do not enable non-trivial transitions?}
\end{itemize}
\end{theorem}

\sj{check correctness for all cases!}

\paragraph{Proof Ideas.}
We explain our proof ideas as modifications of the original proofs by Au{\ss}erlechner et al.~\cite{AusserlechnerJK15}, for the results given in the second results column of Table~\ref{tab:cutoffs-disj}.

In the original proofs corresponding to the first four items, to simulate a given run of an arbitrarily large system, up to $|B|$ processes of the cutoff system are moved into the states that appear in the original run, in the same order. This ensures that all transitions will also be enabled in the cutoff system. Based on our knowledge about guards, we guarantee the same effect by moving into one \emph{representative} state per guard. In this way, we can replace (one occurrence of) $|B|$ by $|G|$ in the cutoff. 

By a similar argument, in the first item we can also replace $|B|$ by $|B|_G$ (this does not work for the other items since additional processes may be needed to ensure fairness or preserve the deadlock).

For local deadlocks, there is an additional construction in the proofs where a process in the cutoff system has to move into some state and then leave it again, because otherwise the deadlock would not be possible. We compute $m$ as an upper bound for the number of states for which this is necessary, which replaces an occurrence of $|B|-1$ in the cutoff.

Finally, for global deadlocks the original proof distinguishes between states in $\mN$ and other states. To construct a simulating run in the cutoff system, for each state in $\mN$ that appears in the deadlocked global state it uses one process that exactly mimics the behavior of one process that moved there in the original run. For the processes that do deadlock in local states that are not in $\mN$, a construction similar to the local deadlocks is needed, moving processes into all states that are visited in the original run, and possibly moving them out of these states again if they are not part of the deadlock. Our improvement concerns only the first set of processes: we compute $\mN^*$ in order to find out how many states from $\mN$ can appear together in a global deadlock. Then, we can replace one occurrence of $|B|-1$ with $|\mN^*|$ in the cutoff.
\qed

\paragraph{Remark.} To compute $\mN^*$ exactly, we need to find the smallest set of states in $\mN$ that do not satisfy the additional condition. This amounts to finding the minimum vertex cover (MVC) for the graph with vertices from $\mN$ and edges from $q_i$ to $q_j$ if $q_i \in \Enable_{q_j}$.\sj{right?} This problem is itself $NP$-hard. This effort is justified since model checking complexity is in general exponential in the number of components. On the other hand, the MVC can be approximated in $PTIME$ such that at least half of the unnecessary nodes are removed.

\begin{table}
\caption{Cutoff Results for Disjunctive Systems}
\label{tab:cutoffs-disj}
\centering
\def\arraystretch{1.5}
\resizebox{\textwidth}{!}{
{\sffamily \small
\begin{tabular}{@{}llccc@{}}
\toprule
&  & EK~\cite{Emerson00} & AJK~\cite{AJK16} & our work \\ 
\midrule
$k$-indexed \LTLmX 						& non-fair~~ 	& ~$|B|+ k+1$~ 				& ~$|B|+k+1$~ 	& ~$|B|_G+k+1$ and $|G|+k+1$~ \\
$k$-indexed \LTLmX 						& fair 			& -	& $2|B|+k-1$ 	& $|B|+|G|+k$ \\
Local Deadlock 	& non-fair 	& -	& $|B|+2$ 	& $m + |G|+1$, with $m<|B|$\\
Local Deadlock 	& fair 			& -	& $2|B|-1$ 	& $|B| + |G|$\\
Global Deadlock & 					& -	& $2|B|-1$ 	& $|B| + |\mN^*|$ with $|\mN^*|<|B|$\\
\bottomrule
\end{tabular}
}
}
\end{table}

\section{Better Cutoffs for Conjunctive Systems}
\label{sec:conj}
In  this section, we state our new cutoff results for conjunctive systems, and compare them to the previously known results in Table~\ref{tab:cutoffs-conj}.
Full proofs can be found in Appendix~\ref{sec:proofs-conj}.

For conjunctive systems, the cutoffs for \LTLmX properties cannot be improved. We give improved cutoffs for global deadlock detection in general, and for local deadlock detection for the restricted case of $1$-conjunctive systems. After that, we explain why local deadlock detection in general is hard, and identify a number of cases where we can solve the problem even for systems that are not $1$-conjunctive.

To state our theorems for conjunctive systems, we define the following for a given conjunctive system $(A, B)^{(1,n)}$:

 We say that $D\subseteq (Q_A \cupdot Q_B)$ is a \emph{deadset} of $q \in (Q_A \cupdot Q_B)$  if $\forall (q,\sigma,g,q') \in \trans: \exists q'' \in D: q'' \notin g$ and  $\forall q'' \in D$ $ \exists (q,\sigma,g,q') \in {\trans}: q'' \not\in g$, and $D$ contains at most one state from $Q_A$. 

 For a given $q$, $dead^\land_q$ is the set of all deadsets of $q$: $dead^\land_q = \{ D \subseteq (Q_A \cupdot Q_B) \mid D \textrm{ is a deadset of } q\}$.

If $dead^\land_q = \emptyset$, then we say $q$ is \emph{free}. If a state $q$ does not appear in $dead^\land_{q'}$ for any $q' \in Q_A \cupdot Q_B$, then we say $q$ is \emph{non-blocking}. If a state $q$ does not appear in $dead^\land_q$, then we say $q$ is \emph{not self-blocking}. 

\begin{theorem}[Conjunctive Cutoff Theorem]
\label{thm:conj}
For conjunctive systems and process templates $A,B$:
\begin{itemize}
\item let 
\begin{itemize}
\item $k_1 = |D_1|$, where $D_1 \subseteq Q_B$ is the set of free states in $B$, 
\item $k_2 = |D_2 \setminus D_1|$, where $D_2 \subseteq Q_B$ is the set of non-blocking states in $B$, and 
\item $k_3 = | D_3 \setminus (D_1 \cup D_2)|$, where $D_3 \subseteq Q_B$ is the set of not self-blocking states in $B$.
\end{itemize}
Then $2|B|-2k_1-2k_2-k_3$ is a cutoff for global deadlock detection.
\item if process template $U$ is $1$-conjunctive, then
\begin{itemize}
\item $|G_U| + 2$ is a cutoff for local deadlock detection in a $U$-process and non-fair executions,
\item $2|G_U| + 1$ is a cutoff for local deadlock detection in an initializing $U$-process and fair executions.
\end{itemize}
\end{itemize}
\end{theorem}

\paragraph{Proof Ideas.}
Again, we explain our proof ideas as modifications of the original proofs by Au{\ss}erlechner et al.~\cite{AusserlechnerJK15}, in this case for the results given in the second results column of Table~\ref{tab:cutoffs-conj}.

In order to simulate a global deadlock of a large system in the cutoff system, the original proof uses up to $2$ processes that move into each of the states --- except for the initial state, which is assumed to be included in every conjunctive guard, and therefore cannot contribute to a deadlock. A generalization of this idea is our notion of non-blocking states, which can further reduce the cutoff. In part, this also applies to states that are not self-blocking: for these, we need at most $1$ copy, since the second copy can only be useful for blocking transitions from the same state. Finally, also states that are free can never contribute to a deadlock, since they are never deadlocked themselves.

Regarding local deadlocks in $1$-conjunctive systems, the idea is similar to the basic idea described in the proof of Theorem~\ref{thm:disj}: where the original proof needs up to one copy of every state (except \init) to ensure that the deadlock is preserved, we need at most one copy for every guard in the template. Therefore, we can replace $|B|-1$ by $|G_U|$ in the cutoff. In the fair case, by a similar argument we can even replace $2|B|-2$ by $2|G_U|+1$.
\qed

\sj{add approach that computes all deadsets and checks their reachability separately, like having a separate cutoff for each deadset?}

\begin{table}
\caption{Cutoff Results for Conjunctive Systems}
\label{tab:cutoffs-conj}
\centering
\def\arraystretch{1.3}
{\sffamily \small
\begin{tabular}{@{}llccc@{}}
\toprule
&  & EK~\cite{Emerson00} & AJK~\cite{AJK16} & our work \\ 
\midrule
$k$-indexed \LTLmX 						& non-fair 	& $k+1$ 					& $k+1$ 				& unchanged \\
$k$-indexed \LTLmX 						& fair 			& - & $k+1$ 				& unchanged \\
Local Deadlock 	& non-fair 	& - & $|B|+1^*$ 	& $|G_U|+2^*$\\
Local Deadlock 	& fair 			& - & $2|B|-2^*$ 	& $2|G_U|+1^*$\\
Global Deadlock & 					& ~$2|B|+1$~ 					& ~$2|B|-2$~ 		& ~$2|B|-2k_1-2k_2-k_3$~\\
\bottomrule
\end{tabular}

$^*:$ systems have to be $1$-conjunctive; in fair case, they additionally have to be initializing;\\
$k_1$: number of free states;\\
$k_2$: number of non-blocking states (that are not free);\\
$k_3$: number of not self-blocking states (that are not free or non-blocking)
}
\end{table}

\paragraph*{Local Deadlock Detection: Beyond 1-conjunctive Systems}
While we improve on the local deadlock detection cutoff for conjunctive systems in some cases, the results above still have the same restriction as in Au{\ss}erlechner et al.~\cite{AJK16}: process template $B$ has to be $1$-conjunctive. The reason for this restriction is that when going beyond $1$-conjunctive systems, the local deadlock detection cutoff (even without considering fairness) can be shown to grow at least quadratic in the number of states or guards, and it becomes very hard to determine a cutoff.

To analyze these cases, define the following:
A sequence of states $q_1 \ldots q_n$ is \emph{connected} if $\forall q_i \in \{q_1, \ldots ,q_n\}: $  $\exists (q_i,\sigma,g,q_{i+1}) \in {\trans}$.
A \emph{cycle} is a connected sequence of states $q\, q_1 \ldots q_n\, q$ such that $\forall q_i,q_j \in \{q_1, \ldots, q_n\}:$ $q_i \neq q_j$. We denote such a cycle by $C_q$. (By abuse of notation, $C_q$ is also used for the set of states on $C_q$.) We denote the set of guards of the transitions on $C_q$ as $G_{C_q}$. 
A \emph{cycle} $C_q$ is called \emph{free} if $\forall p \in C_q \setminus q$ $\forall g \in G_{C_q}:$ $p \in g$. We denote such a cycle by $C_q^{free}$.

\begin{example}
\label{ex:quadratic-cutoff}
If we consider the process template in Figure~\ref{fig:quadratic-cutoff} without the parts in blue, then it exhibits a local deadlock in state $q_l$ for $9$ processes, but not for $8$ processes: one process has to move to $q_l$, and for each cycle that starts and ends in states $a,b,c,d$, we need $2$ processes that move along the cycle to keep all guards of $q_l$ covered at all times. Intuitively, one copy per cycle has to be in the state of interest, or ready to enter it, and the other copy is traveling on the cycle, waiting until the guards are satisfied.

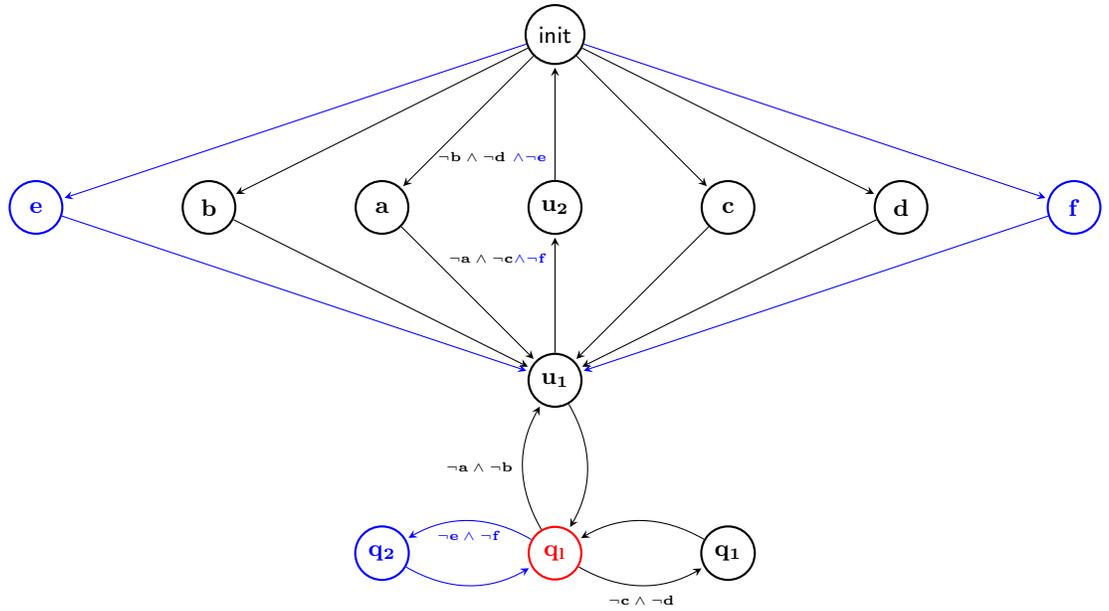
\begin{figure}
\begin{tikzpicture}[node distance=2.3cm,>=stealth,auto]
  \tikzstyle{state}=[circle,thick,draw=black,minimum size=7mm]
  \tikzstyle{qlstate}=[circle,thick,draw=red,color=red,minimum size=7mm]
  \tikzstyle{dstate}=[circle,thick,color=blue,draw=blue,minimum size=7mm]

  \begin{scope}
    
    \node [state] (init) {$\init$};
    
    \node [state] (u1) [below of=init] {$\mathbf{u_2}$}
    edge [post] node[pos = 0.2pt] {\tiny{$\mathbf{\neg b \land \neg d}$ {\color{blue} $\mathbf{\land \neg e}$}}} (init); 
    
    \node [state] (u2) [below of=u1] {$\mathbf{u_1}$}
    edge [post] node[pos = 0.8] {\tiny{$\mathbf{\neg a \land \neg c}${\color{blue} $\mathbf{\land \neg f}$}}} (u1); 
    
    \node [qlstate] (ql) [below of=u2] {$\mathbf{q_l}$}
    edge [pre, bend right = 30] (u2)
    edge [post, bend left = 30] node[left = 0.5pt] {\tiny{$\mathbf{\neg a \land \neg b}$}} (u2);
    
    \node [state] (u3) [right of=ql] {$\mathbf{q_1}$}
    edge [post, bend right = 30] (ql)
    edge [pre, bend left = 30] node[below = 0.5pt] {\tiny{$\mathbf{\neg c \land \neg d}$}} (ql);
    
    \node [dstate] (u4) [left of=ql] {$\mathbf{q_2}$}
    edge [post, blue, bend right = 30] (ql)
    edge [pre, blue, bend left = 30] node[below = 0.5pt] {\tiny{$\mathbf{\neg e \land \neg f}$}} (ql);
    
    \node [state] (a) [left of=u1] {$\mathbf{a}$}     
     edge [pre] (init)
     edge [post] (u2);
     
     \node [state] (b) [left of=a] {$\mathbf{b}$}     
     edge [pre] (init)
     edge [post] (u2);
     
     \node [dstate] (e) [left of=b] {$\mathbf{e}$}     
     edge [pre, blue] (init)
     edge [post, blue] (u2);
     
     \node [state] (c) [right of=u1] {$\mathbf{c}$}     
     edge [pre] (init)
     edge [post] (u2);
     
     \node [state] (d) [right of=c] {$\mathbf{d}$}     
     edge [pre] (init)
     edge [post] (u2);
     
     \node [dstate] (f) [right of=d] {$\mathbf{f}$}     
     edge [pre, blue] (init)
     edge [post, blue] (u2);

  \end{scope}
\end{tikzpicture}
\caption{Process Template with Quadratic Cutoff for Local Deadlocks}
\label{fig:quadratic-cutoff}
\end{figure}

Now, consider the modified template (as depicted in blue in Figure~\ref{fig:quadratic-cutoff}) where we i) add two states $e,f$ in a similar way as $a,b,c,d$, ii) add a new state connected to $q_l$ with guard $\neg e \land \neg f$, and iii) change the guards in the sequence from $u_1$ to $\init$ to $\neg a \land \neg c \land \neg e$ and $\neg b \land \neg d \land \neg f$, respectively. Then we have $6$ cycles that need $2$ processes each, and we need $13$ processes to reach a local deadlock in $q_l$.

Moreover, consider the modified template where we increase the length of the sequence from $u_1$ to $\init$ by adding additional states $u_3$ (which is connected to $u_2$ instead of $\init$) and $u_4$ (which is connected to $u_3$ and $\init$ with transitions that have the same guards as those from $u_1$ to $u_2$ and from $u_2$ to $u_3$, respectively). Then, for every cycle we need $3$ processes instead of $2$, as otherwise they cannot traverse the cycle fast enough to ensure that the local deadlock is preserved infinitely long. That is, the template with both modifications now needs $19$ processes to reach a local deadlock. Observe that by increasing the height of the template, we increase the necessary number of states without increasing the number of different guards.

Moreover, when increasing both the width and height of the template, we observe that the number of processes that are necessary for a local deadlock increases quadratically with the size of the template.
\end{example}

This example leads us to the following result.

\begin{theorem}
For conjunctive systems, a cutoff for local deadlock detection must grow at least quadratically in the number of states. Furthermore, it cannot be bounded by the number of guards at all.
\end{theorem}

\paragraph{Proof Idea.}
For a system that does exhibit a local deadlock for some size $n$, but not for $n-1$, the cutoff cannot be smaller than $n$. Thus, the example shows that a cutoff for local deadlock detection in general is independent of the number of guards, and must grow at least quadratic in the size of the template.  \qed

\medskip
Cutoffs that can in the best case be bounded by $|B|^2$ will not be very useful in practice. Therefore, instead of solving the general problem we identify in the following a number of cases where the cutoff remains small (i.e., linear in the number of states or guards).

When comparing the proof of the second item of Theorem~\ref{thm:conj} to the example above, we note that the reason that the cutoff in Theorem~\ref{thm:conj} does not apply is the following: while in $1$-conjunctive systems every state has a unique deadset, in the general case every state may have many deadsets, and the structure of the process template may require infinitely many alternations between different deadsets to preserve the local deadlock. Moreover, as shown in the example, the number of processes needed to alternate between deadsets may increase with the size of the template, even if the set of guards (and thus, the number of different deadsets) remains the same.

We say that a locally deadlocked run is \emph{alternation-free} if it does not alternate infinitely often between different deadsets.
In the following, we will first show that for certain systems with alternation-free local deadlocks, the cutoff for $1$-conjunctive systems applies.
After that, we consider a (still restricted) class of systems that does not have alternation-free local deadlocks, and give a local deadlock detection cutoff for this class.

\paragraph{Systems with Alternation-Free Local Deadlocks.}

\sj{idea: prove that $1$-conjunctive cutoff, per state, applies iff for every local deadlock there exists an alternation-free local deadlock; then give sufficient conditions for this; does it work if different states have different properties?} 

%

To identify systems with alternation-free deadlocks, we need some additional definitions.

We say that a conjunctive process template $U$ is \emph{effectively $1$-conjunctive} if every $q \in Q_U$ is either $1$-conjunctive or free.

A \emph{lasso} $lo$ is a connected sequence of states $q_0 \ldots q_i \ldots q_n$ such that $q_0$ is an initial state, $q_i = q_n$, and $q_i \ldots q_n$ is a cycle.
We denote by $G_{lo}$ the set of guards of the transitions on $lo$.
We say that a conjunctive process template $U$ is \emph{freely traversable} if for every non-free state $q \in Q_U$, and every set of states $\{q_1, \ldots , q_n\}$ that disables the $n$-conjunctive guards with $n>1$ in transitions from $q$, there exists a lasso $lo$ that is free of $\neg q$, free of all $\neg q_i$, and free of all $1$-conjunctive guards in transitions from $q$.

Intuitively, in a freely traversable process template there is always an infinite local run that can start from \init when a single other process is already in a local deadlock. The example process in Section~\ref{sec:intro} is not freely traversable, since there is a lasso that is free of $\neg tw$ and $\neg r$, but no lasso that is free of $\neg tw$ and $\neg w$.

We say that a conjunctive process template $U$ is \emph{alternation-free} if one of the following holds:
\begin{itemize}
\item for every non-free state $q \in Q_U$, and every set of states $D= \{q_1, \ldots , q_n\}$ that disables the $n$-conjunctive guards with $n>1$ in transitions from $q$, there is at most one $q_i$ for which the following does not hold:
$$\textrm{for all cycles } C_{q_i} = q_i \ldots q_i \in U: C_{q_i} \cap (C_q \cup \neg q) \neq \emptyset$$
\item for every non-free state $q \in Q_U$, and every $n$-conjunctive guard $g = \neg q_1 \land \ldots \land q_n$ with $n>1$, $G_q \cap \{\neg q_1, \ldots, \neg q_n\} \neq \emptyset$. 
\end{itemize}

Intuitively, in an alternation-free process template there can never be an infinite alternation between different deadsets of a single locally deadlocked process (without releasing the deadlock). The process template from Section~\ref{sec:intro} is alternation-free, since: i) $tw$ is the only non-free state with guards that are not $1$-conjunctive, ii) $\{w,r\}$ is the set of states that disables the only guard that is not $1$-conjunctive, and iii) all cycles that contain $w$ also contain a guard that is in $G_{tw}$ (since all these cycles move through $tw$).

\begin{obs}
If a process template $U$ is either effectively $1$-conjunctive, freely traversable, or alternation-free, then for every locally deadlocked run there exists a locally deadlocked run that is alternation-free.
\end{obs}

\begin{theorem}[Local Deadlock Detection in Conjunctive Systems]
\label{thm:conj-ld}
For conjunctive systems and process templates $A,B$, for $U\in \{A,B\}$ the respective cutoff for local deadlock detection in $1$-conjunctive systems applies in the following cases:
	\begin{itemize}
	\item for non-fair executions if $U$ is effectively $1$-conjunctive, freely traversable, or alternation-free
	\item for unconditionally fair executions if $U$ is effectively $1$-conjunctive or alternation-free.
\end{itemize}
\end{theorem}

\paragraph{Proof Ideas.}
The statement follows from the observation above, and from the proof of Theorem~\ref{thm:conj}. Only the notion of freely traversable process templates is not compatible with the proof for local deadlocks under fairness.
\qed

\paragraph{Systems without Alternation-free Local Deadlocks.}
To demonstrate the complexity of the problem in general, let us analyze a non-trivial, but still strongly restricted case where alternation between deadsets may be necessary. Consider a system where all non-trivial guards are $1$-conjunctive, except for a single $2$-conjunctive guard $g^2 = \neg a \land \neg b$ that is used in a single transition from state $q_l$. To simplify the analysis, assume that the process template has unique cycles $C_a$ and $C_b$, i.e., no other cycles pass through $a$ or $b$. Assume that both cycles are free of $1$-conjunctive guards that are necessary to deadlock $q_l$, and free of $\neg q_l$ (otherwise, the template would be alternation-free).

To state the cutoff result, define the following:
A \emph{segment} $Sg_{a-b}$ is a connected sequence of states $q_i \ldots q_j$ where:
\begin{itemize}
	\item $q_i$ has an incoming transition with guard $\neg a$
	\item $q_j$ has an outgoing transition with guard $\neg b$
	\item $\forall q_m \in Sg_{a-b}$ $\exists (q_m,\sigma, g, q_{m+1}) \in \delta:$ if $q_{m+1} \in Sg_{a-b}$ then $ b \in g$
\end{itemize}
For a cycle $C_q$, we denote by $|Sg_{a-b}|_{C_q}$ the total number of segments $Sg_{a-b}$ on $C_q$

\begin{theorem}
\label{thm:1-2-conj}
For a system with process templates $A,B$ and the restrictions described above, let $n_a = max(|Sg_{a-b}|_{C_a}  ,|Sg_{b-a}|_{C_a})$ and $n_b = max(|Sg_{a-b}|_{C_b}  , |Sg_{b-a}|_{C_b}).$
Then: 
 $$(A,B)^{(1,n)} \textrm{ has a local deadlock in $q_l$} \implies (A,B)^{(1,|G_B| + n_a + n_b + 5)} \textrm{ has a local deadlock in $q_l$}.$$ 
\end{theorem}

That is, already for this restricted class of systems, the available proof methods only give us a cutoff that increases with the number of segments $Sg_{a-b}$ and $Sg_{b-a}$ on the cycles. For systems with multiple $n$-conjunctive guards, both the complexity of the analysis and the size of the cutoff grow quickly (and Example~\ref{ex:quadratic-cutoff} shows that this may indeed be necessary).
\section{Verification of the Reader-Writer Example}

We consider again the reader-writer example from Section~\ref{sec:intro}, and show how our new results allow us to check correctness, find a bug, and check a fixed version.

With our results, we can for the first time check this liveness property in a meaningful way, i.e., under the assumption of fair scheduling. Since the process template is alternation-free, by Theorems~\ref{thm:conj} and \ref{thm:conj-ld} the local deadlock detection cutoff for the system is $2|G_B|+1=5$. 
Moreover, compared to previous results we reduce the cutoff for global deadlock detection by recognizing that $k_1=3$ states can never be deadlocked, and $k_2=2$ additional states never appear in any guard. This reduces the cutoff to $2|B|-2k_1-2k_2=10-6-4=0$, i.e., we detect that there can be no global deadlocks by analyzing only a single process template.

However, checking the system for local deadlocks shows that a local deadlock is possible: a process may forever be stuck in $tw$ if the other processes move in a loop $(\init, tr, r)^\omega$ (and always at least one process is in $r$).
To fix this, we can add an additional guard $\neg tw$ to the 
\begin{wrapfigure}{r}{0.4\linewidth}
\vspace{-8pt}
\scalebox{0.95}{
	\begin{tikzpicture}[node distance=2cm,>=stealth,auto]
  \tikzstyle{state}=[circle,thick,draw=black,minimum size=7mm]

  \begin{scope}
    
    \node [state] (init) {$\init$};   
    
    \node [state] (r) [left of=init] {$\mathbf{r}$}     
     edge [post] (init);
     
    \node [state] (tr) [below of=r] {$\mathbf{tr}$}    
     edge [pre] node[right =0.5pt] {\tiny{$\mathbf{\neg tw}$}} (init) 
     edge [post] node[left = 0.5pt] {\tiny{$\mathbf{\neg w}$}} (r);
     
    \node [state] (w) [right of=init] {$\mathbf{w}$}     
     edge [post] (init);    
     
    \node [state] (tw) [below of=w] {$\mathbf{tw}$}     
     edge [pre] (init)
     edge [post] node[right =0.5pt] {\tiny{$\mathbf{\neg w \land \neg r}$}}(w);

  \end{scope}
  \end{tikzpicture}
}
\label{fig:reader-writer}
\vspace{-20pt}
\end{wrapfigure}
transition from $\init$ to $tr$, as shown in the 
process template to the right. 
For the resulting system, our results give a local deadlock detection cutoff of $2|G_B|+1=7$, and a global deadlock detection cutoff of $2|B|-2k_1-2k_2-k_3=10-6-2-1=1$ (where $k_3$ is the number of states that do appear in guards and could be deadlocked themselves, but do not have a transition that is blocked by another process in the same state).
\section{More Disjunctive Systems and More Specifications}

We show two further extensions of the class of problems for which cutoffs are available:
\begin{enumerate}
\item systems where transitions are guarded with a conjunction of disjunctive guards
\item two important classes of specifications that cannot be expressed in prenex indexed temporal logic.
\end{enumerate} 

\subsection{Systems with Conjunctions of Disjunctive Guards}

We consider systems where a transition can be guarded by a set of sets of states, interpreted as a conjunction of disjunctive guards. I.e., a guard $\{D_1,\ldots,D_n\}$ is satisfied in a given global state if for all $i=1,\ldots,n$, there exists another process in a state $D_i$. 

We observe that for this class of systems, most of the original proof ideas still work. For results that depend on the number of guards, we have to count the number of different conjuncts in guards. 

\begin{theorem}
For systems with conjunctions of disjunctive guards, cutoff results for disjunctive systems that do not depend on the number of guards still hold (first and second column of results in Table~\ref{tab:cutoffs-disj}).

Cutoff results that depend on the number of guards (last column of Table~\ref{tab:cutoffs-disj}) hold if we consider the number of conjuncts in guards instead. For results that additionally refer to some measure of the sets of enabling states ($m$ and $|\mN^*|$, respectively), we obtain a valid cutoff for systems with conjunctions of disjunctive guards if we replace this measure by $|B|-1$.
\end{theorem}

\paragraph{Proof Ideas.}
The cutoff results that are independent of the number of guards still hold since all of the original proof constructions still work. To simulate a run $x$ of a large system in a run $y$ the cutoff system, one task is to make sure that all necessary transitions are enabled in the cutoff system. To this end, the original construction of $y$ moves one process into each state that appears in $x$, as soon as possible. This ensures that if we only want to enter states that appear in the original run, disjunctive guards of all necessary transitions will be satisfied. However, the same holds for transitions with conjunctions of disjunctive guards --- if the set of states that appear in the other processes is the same at a given time, then the same conjunctions of disjunctive guards will be satisfied.

By a similar argument, we can always move out of a state if necessary for the construction, and deadlocks are preserved in the same way as for disjunctive systems.

For cutoffs that depend on the number of guards, transitions with conjunctions of disjunctive guards require us to use one representative for each conjunct in a guard, in the construction explained in the proof idea of Theorem~\ref{thm:disj}.

Finally, the reductions of the cutoff based on the analysis of states that can or cannot appear together in a deadlock do not work in these extended systems, and we have to replace $m$ and $|\mN^*|$ by $|B|-1$ in the cutoffs. The reason is that $\Enable_q$ is now not a set of states anymore, but a set of sets of states. A more detailed analysis based on this observation may be possible, but is left open for now.
\qed

\subsection{Simultaneous Reachability of Target States}

An important class of properties for parameterized systems asks for the reachability of a global state where all processes of type $B$ are in a given local state $q$ (compare Delzanno et al.~\cite{DelzannoSZ10}). This can be written in indexed \LTLmX as $\eventually \forall i. q_i$, but is not expressible in the fragment where index quantifiers have to be in prenex form. We denote this class of specifications as \textsc{Target}. 
Similarly, repeated reachability of $q$ by all states simultaneously can be written $\GF \forall i. q_i$, and is also not expressible in prenex form. We denote this class of specifications as \textsc{Repeat-Target}.

\begin{theorem}[Disjunctive \textsc{Target} and \textsc{Repeat-Target}]
For disjunctive systems: $|B|$ is a cutoff for checking \textsc{Target} and \textsc{Repeat-Target}.
\end{theorem}

\paragraph{Proof Ideas.}
We can simulate a run $x$ in a large system where all processes are in $q$ at time $m$ in the cutoff system by first moving one process into each state that appears in $x$ before $m$, in the same order as in $x$. To make all processes reach $q$, we move them out of their respective states in the same order as they have moved out of them in $x$. For this construction, we need at most $|B|$ processes.

If in $x$ the processes reach are repeatedly in $q$ at the same time, then we can simulate this also in the cutoff system: if $m'>m$ is a point in time where this happens again, then we use the same construction as above, except that we consider all states that are visited between $m$ and $m'$, and we move to these states from $q$ instead from $\init$. The correctness argument is the same, however.

Finally, if the run with \textsc{Repeat-Target} should also be fair, then we do not simply select any $m'$ with the property above, but we choose it such that all processes move between $m$ and $m'$. If the original run $x$ is fair, then such an $m'$ must exist.
\qed

\paragraph{\textsc{Target} and \textsc{Repeat-Target} in Conjunctive Systems.}
For conjunctive systems, obtaining cutoffs for \textsc{Target} and \textsc{Repeat-Target} is hard, for similar reasons as obtaining a cutoff for local deadlock detection is hard in general (see Section~\ref{sec:conj}). We leave this as an open question.
\section{Conclusion}

We have shown that better cutoffs for guarded protocols can be obtained by analyzing properties of the process templates, in particular the number and form of transition guards. We have further shown that cutoff results for disjunctive systems can be extended to a new class of systems with conjunctions of disjunctive guards, and to specifications \textsc{Target} and \textsc{Repeat-Target}, that have not been considered for guarded protocols before.

For conjunctive systems, previous works have treated local deadlock detection only for the restricted case of systems with $1$-conjunctive guards. We have considered the general case, and have shown that it is very difficult --- the cutoffs grow independently of the number of guards, and at least quadratically in the size of the process template. To circumvent this worst-case behavior, we have identified a number of conditions under which a small cutoff can be obtained even for systems that are not $1$-conjunctive.

By providing cutoffs for systems and specifications that were previously not known to have cutoffs or to be decidable, we have in particular proved decidability of the respective problems.

Our work is inspired by applications in
parameterized synthesis~\cite{Jacobs14}, where the goal is to automatically
\emph{construct} process templates such that a given specification is satisfied in
systems with an arbitrary number of components. In this setting, deadlock detection and expressive specifications are particularly important, since \emph{all} relevant properties of the system have to be specified, in contrast to verification, where a partial specification may be acceptable. The results of this paper can be seen as a continuation of our research on efficient parameterized synthesis, orthogonal to the approaches like modular application of cutoffs presented in earlier work~\cite{Khalimov13}.

Besides making verification and synthesis more efficient through smaller cutoffs, our results can also be used to guide synthesis algorithms towards ``simple'' implementations, that have additional benefits such as being easier to understand, verify, and maintain (by humans and machine alike). This approach has been used by others before: \emph{bounded synthesis}~\cite{FinkbeinerS13} prefers implementations with a small number of states, \emph{bounded cycle synthesis}~\cite{FinkbeinerK16} prefers implementations with a small number of cycles.
Investigating the applications of our results in parameterized synthesis is one of our goals in future work.

\paragraph{Acknowledgements.}
We thank Ayrat Khalimov for fruitful discussions on guarded protocols, and Martin Zimmermann for suggestions regarding a draft of this work.
\bibliographystyle{plainurl}
\bibliography{paper,local,references,crossrefs}

\begin{thebibliography}{10}

\bibitem{AJKR14}
B.~Aminof, S.~Jacobs, A.~Khalimov, and S.~Rubin.
\newblock Parameterized model checking of token-passing systems.
\newblock In {\em VMCAI}, volume 8318 of {\em LNCS}, pages 262--281. Springer,
  2014.

\bibitem{AusserlechnerJK15}
Simon Au{\ss}erlechner, Swen Jacobs, and Ayrat Khalimov.
\newblock Tight cutoffs for guarded protocols with fairness.
\newblock {\em CoRR}, abs/1505.03273, 2015.
\newblock Extended version with full proofs.
\newblock URL: \url{http://arxiv.org/abs/1505.03273}.

\bibitem{AJK16}
Simon Au{\ss}erlechner, Swen Jacobs, and Ayrat Khalimov.
\newblock Tight cutoffs for guarded protocols with fairness.
\newblock In {\em {VMCAI}}, volume 9583 of {\em LNCS}, pages 476--494.
  Springer, 2016.
\newblock \href {http://dx.doi.org/10.1007/978-3-662-49122-5_23}
  {\path{doi:10.1007/978-3-662-49122-5_23}}.

\bibitem{PrinciplesMC}
Christel Baier and Joost-Pieter Katoen.
\newblock {\em Principles of model checking}, volume 26202649.
\newblock MIT press Cambridge, 2008.

\bibitem{BloemETAL15}
Roderick Bloem, Swen Jacobs, Ayrat Khalimov, Igor Konnov, Sasha Rubin, Helmut
  Veith, and Josef Widder.
\newblock {\em Decidability of Parameterized Verification}.
\newblock Synthesis Lectures on Distributed Computing Theory. Morgan {\&}
  Claypool Publishers, 2015.
\newblock \href {http://dx.doi.org/10.2200/S00658ED1V01Y201508DCT013}
  {\path{doi:10.2200/S00658ED1V01Y201508DCT013}}.

\bibitem{Bouajjani00}
A.~Bouajjani, B.~Jonsson, M.~Nilsson, and T.~Touili.
\newblock Regular model checking.
\newblock In {\em CAV}, volume 1855 of {\em LNCS}, pages 403--418. Springer,
  2000.
\newblock \href {http://dx.doi.org/10.1007/10722167_31}
  {\path{doi:10.1007/10722167_31}}.

\bibitem{Clarke08}
E.~M. Clarke, M.~Talapur, and H.~Veith.
\newblock Proving ptolemy right: The environment abstraction framework for
  model checking concurrent systems.
\newblock In {\em TACAS}, volume 4963 of {\em LNCS}, pages 33--47. Springer,
  2008.

\bibitem{Clarke04c}
E.~M. Clarke, M.~Talupur, T.~Touili, and H.~Veith.
\newblock Verification by network decomposition.
\newblock In {\em CONCUR}, volume 3170 of {\em LNCS}, pages 276--291. Springer,
  2004.

\bibitem{DelzannoSZ10}
Giorgio Delzanno, Arnaud Sangnier, and Gianluigi Zavattaro.
\newblock Parameterized verification of ad hoc networks.
\newblock In {\em {CONCUR}}, volume 6269 of {\em LNCS}, pages 313--327.
  Springer, 2010.
\newblock \href {http://dx.doi.org/10.1007/978-3-642-15375-4_22}
  {\path{doi:10.1007/978-3-642-15375-4_22}}.

\bibitem{Emerson00}
E.~A. Emerson and V.~Kahlon.
\newblock Reducing model checking of the many to the few.
\newblock In {\em CADE}, volume 1831 of {\em LNCS}, pages 236--254. Springer,
  2000.
\newblock \href {http://dx.doi.org/10.1007/10721959_19}
  {\path{doi:10.1007/10721959_19}}.

\bibitem{EmersonK03}
E.~A. Emerson and V.~Kahlon.
\newblock Model checking guarded protocols.
\newblock In {\em LICS}, pages 361--370. IEEE Computer Society, 2003.
\newblock \href {http://dx.doi.org/10.1109/LICS.2003.1210076}
  {\path{doi:10.1109/LICS.2003.1210076}}.

\bibitem{Emerso03}
E.~A. Emerson and K.~S. Namjoshi.
\newblock On reasoning about rings.
\newblock {\em Foundations of Computer Science}, 14:527--549, 2003.

\bibitem{EsparzaFM99}
J.~Esparza, A.~Finkel, and R.~Mayr.
\newblock On the verification of broadcast protocols.
\newblock In {\em LICS}, pages 352--359. {IEEE} Computer Society, 1999.
\newblock \href {http://dx.doi.org/10.1109/LICS.1999.782630}
  {\path{doi:10.1109/LICS.1999.782630}}.

\bibitem{Esparza14}
Javier Esparza.
\newblock Keeping a crowd safe: On the complexity of parameterized verification
  (invited talk).
\newblock In {\em {STACS}}, volume~25 of {\em LIPIcs}, pages 1--10. Schloss
  Dagstuhl - Leibniz-Zentrum fuer Informatik, 2014.
\newblock \href {http://dx.doi.org/10.4230/LIPIcs.STACS.2014.1}
  {\path{doi:10.4230/LIPIcs.STACS.2014.1}}.

\bibitem{FinkbeinerS13}
B.~Finkbeiner and S.~Schewe.
\newblock Bounded synthesis.
\newblock {\em STTT}, 15(5-6):519--539, 2013.
\newblock \href {http://dx.doi.org/10.1007/s10009-012-0228-z}
  {\path{doi:10.1007/s10009-012-0228-z}}.

\bibitem{FinkbeinerK16}
Bernd Finkbeiner and Felix Klein.
\newblock Bounded cycle synthesis.
\newblock In {\em {CAV} {(1)}}, volume 9779 of {\em LNCS}, pages 118--135.
  Springer, 2016.
\newblock \href {http://dx.doi.org/10.1007/978-3-319-41528-4_7}
  {\path{doi:10.1007/978-3-319-41528-4_7}}.

\bibitem{German92}
S.~M. German and A.~P. Sistla.
\newblock Reasoning about systems with many processes.
\newblock {\em J. ACM}, 39(3):675--735, 1992.

\bibitem{Jacobs14}
S.~Jacobs and R.~Bloem.
\newblock Parameterized synthesis.
\newblock {\em Logical Methods in Computer Science}, 10:1--29, 2014.

\bibitem{KaiserKW10}
A.~Kaiser, D.~Kroening, and T.~Wahl.
\newblock Dynamic cutoff detection in parameterized concurrent programs.
\newblock In {\em CAV}, volume 6174 of {\em LNCS}, pages 645--659. Springer,
  2010.
\newblock \href {http://dx.doi.org/10.1007/978-3-642-14295-6_55}
  {\path{doi:10.1007/978-3-642-14295-6_55}}.

\bibitem{Khalimov13}
A.~Khalimov, S.~Jacobs, and R.~Bloem.
\newblock Towards efficient parameterized synthesis.
\newblock In {\em VMCAI}, volume 7737 of {\em LNCS}, pages 108--127. Springer,
  2013.

\bibitem{Kurshan95}
R.~P. Kurshan and K.~L. McMillan.
\newblock A structural induction theorem for processes.
\newblock {\em Inf. and Comp.}, 117(1):1--11, 1995.

\bibitem{Suzuki88}
I.~Suzuki.
\newblock Proving properties of a ring of finite state machines.
\newblock {\em Inf. Process. Lett.}, 28(4):213--214, 1988.

\end{thebibliography}
\newpage
\appendix
\section{Appendix: Proofs and Proof Methods for Disjunctive Systems}
\label{sec:proofs-disj}


In this section, we present lemmas and proof methods that allow us to obtain our cutoff results for disjunctive systems. Note that usually we only state a \emph{bounding lemma}, which states that any behavior in a large system can be replicated in the cutoff system. For the opposite direction, we can use existing \emph{monotonicity lemmas} from previous work~\cite{Emerson00,AJK16} (see in particular the full version of Au{\ss}erlechner et al.~\cite{AusserlechnerJK15}).
Also, in many cases we only consider a problem for a copy of template $B$, but not for $A$. The case of $A$ can be obtained by minor modifications of the proofs.

\subsection{Definitions}

Given a run $x=x_0,x_1...$ of a system $(A, B)^{(1,n)}$ and a state $q \in Q_B$, we define the following notation:

\begin{itemize}
\item $\appears_q$ is the set of all moments where at least one copy of $B$ is in state $q$: $\appears_q=\{m \in \Nat \mid \exists i \in [n]: x_m(B_i)=q\}$
\item $f_q$ is the first moment where $q$ appears: $f_q = min(\appears_q)$
\item $\first_q \in [n]$ is the process index with $x_{f_q}(B_{\first_q}) = q$
\item if $\appears_q$ is finite, then $l_q$ is the last moment where $q$ appears: $l_q = max(\appears_q)$
\item $\last_q \in [n]$ is the process index with $x_{l_q}(B_{\last_q}) = q$
\item given a guard $g \in G$, its \emph{representative} is a tuple that contains the state from $g$ that first appears in $x$, and the local run in which this state appears first: a tuple $(x(B_{\first_{q_r}}),q_r)$ is a \emph{representative} for $g$ iff the following holds: $\forall q_i \in g: f_{q_r} 
\leq f_{q_i}$. Note that multiple guards might have the same \emph{representative}.
\item $\occurs_m(q)$ is the number of processes that are in state $q$ at moment $m$: $\occurs_m(q) = |\{B_i \in \mathcal{B}$ $|$ $x_m(B_i) = q\}|$
\end{itemize}

\subsection{$\LTLmX$ Properties, Without Fairness}

In this section, we show how to obtain a cutoff for \LTLmX properties in disjunctive systems without fairness. As mentioned before, we only need to show that a behaviour from a large system can be replicated in the cutoff system. 

\begin{lemma}[Bounding Lemma, \LTLmX, disjunctive, non-fair]
\label{lem:bounding-disj-nf}
For process templates $A,B$ with $G = G_A \cup G_B$ and $n \geq |G|+1$:
$$(A, B)^{(1,n)} \models \pexists h(A,B^{(1)}) \implies (A, B)^{(1,|G|+1)} \models \pexists h(A,B^{(1)})$$
\end{lemma}

\smartpar{Proof.}
Let $x = x_0,x_1,..$ be a run of $\largesys$ that satisfies $h(A,B^{(1)}$. 
We construct a run $y=y_0,y_1...$ of $\cutoffsys$ that satisfies $h(A,B^{(1)}$ as follows:
\begin{enumerate}
\item $y(A) = x(A)$
\item $y(B_1) = x(B_1)$
\item for each $g_j \in G = \{g_1,\ldots,g_k\}$, let $(x(B_{\first_{q_r}}),q_r)$ be the \emph{representative} 
for $g_j$, then $y(B_{j+1}) = x(B_{\first_{q_r}})[1:f_{q_r}](q_r)^w$. In other words, $B_{j+1}
$ imitates $B_{\first_{q_r}}$ until it reaches $q_r$ then it stays in $q_r$ forever. This 
is called \emph{flooding} of a local state $q_r$.
\end{enumerate}

With this construction, it might happen that the run $y$ violates the interleaving semantics 
requirement (i.e., that only one process moves at a time), because it is possible that two 
different guards have the same process \emph{representative} $x(B_i)$. To 
resolve this  problem, we add stuttering steps into local runs whenever two 
or more  processes move at the same time.

The intuition behind the construction is that instead of flooding all states (that appear in the given run), we only flood at most one per guard --- the one that appears first in $x$.

To prove 
correctness, it is enough to prove that at any moment $m$, if a transition $t$ 
for a process is enabled in $x$ then it is enabled in $y$. Now suppose at time $m$
 a transition $t$ is enabled in $x$, then $\exists q \in g_t$ (guard 
of transition $t$) and $\exists p$ such that $x_m(p) = q$, then $q$ enables 
$g_t$ but it is not necessarily a \emph{representative}. In case it is a 
\emph{representative} then by construction $g_t$ is enabled in $y$. 
In case it is not, then either $q \in Q_A$ or $\exists q_r \in g_t$ such that $f_{q_r} \leq 
f_{q}$, and by construction $\exists B_r$ where $y_m(B_r) = q_r
$. In both cases, $g_t$ is enabled in $y$ at time $m$. \qed

\subsection{$\LTLmX$ Properties, With Fairness}
\label{DisjFair}

\begin{lemma}[Bounding Lemma, \LTLmX, disjunctive, fair]
\label{lem:bounding-disj-f}
For process templates $A,B$ with $G = G_A \cup G_B$ and $n \geq |B|+|G|+1$:
\begin{itemize}
\item[] $(A, B)^{(1,n)} \models \pexists (\spec_{uncond} 
\land h(A,B^{(1)})) \implies (A, B)^{(1,|B| + |G| + 1)} \models \pexists (\spec_{uncond} 
\land h(A,B^{(1)}))$
\end{itemize}
\end{lemma}

\smartpar{Proof.}
Let $x=x_0,x_1...$ be a run of $\largesys$ that satisfies $h(A,B^{(1)})$ and unconditional fairness. Given a subset $F \subseteq \mathcal{B}$, define
$$\visited^{inf}_{F} = \{q \in Q_B \mid \appears_q \textrm{ is infinite}\}$$
$$\visited^{fin}_{F} = \{q \in Q_B \mid \appears_q \textrm{ is finite}\}$$
A tuple $(x(B_{\first_{q_r}}),q_r)$ is an \emph{infinite representative} for a guard $g \in G$ if $q_r \in \visited^{inf}_{F}$ and $\forall q_i \in g, q_i \in \visited^{inf}_{F}: f_{q_r} \leq f_{q_i}$.

\smartpar{Construction:} 
We construct a run $y=y_0,y_1...$ of $\cutoffsys$ that satisfies $h(A,B^{(1)})$ and unconditional fairness:
\begin{enumerate}
\item $y(A) = x(A).$
\item $y(B_1) = x(B_1).$
\item to every $q \in \visited^{fin}_{B_2...B_n}$ devote one process $B_{i_q}$
 such that\\
 $y(B_{i_q})=x(B_{\first_q})[1:f_q].(q)^{l_q-f_q}.x(B_{\last_q})[l_q+1:]$\\
This is called \emph{flooding} of state $q$ \emph{with evacuation} into $\visited^{inf}$ (since $B_{\last_q}$ has to move into $\visited^{inf}$ eventually).
\item for each $g \in G_B$, let $(x(B_{\first_{q_r}}),q_r)$ be the infinite 
representative for $g$, and devote two 
processes $B_{g_1}$ and $B_{g_2}$ to $g$, such that $y(B_{g_1})$ and 
$y(B_{g_2})$ imitate $x(B_{\first_{q_r}})$ until the first occurence of $q_r$, then they take 
turns: always one process copies $x(B_{\first_{q_r}})$ while the other 
stutters in $q_r$, and they switch roles every time $x(B_{\first_{q_r}})$ visits $q_r$.
\end{enumerate}
The local runs of the processes devoted to states in $\visited^{fin}_{B_2...B_
n}$ ensure that at any moment the subset of $\visited^{fin}_{B_2...B_n}$ 
that appears in $y$ is a superset of the subset of $\visited^{fin}_{B_2...B_n}$ 
that appears in $x$. Together with the local runs of the processes devoted 
to the guards' infinite representatives, this ensures that any transition 
enabled in $x$ is also enabled in $y$: at any moment, if a state $q_i$ appears in $x$ and either $q_i \in \visited^{fin}_{B_2...B_
n}$ or $q_i$ is an infinite representative of some guard, then it also appears in $y$.

Note that for the finite part we cannot use the guard 
representative, because its ``life span'' may be shorter than we need, and we 
cannot flood it as we need to preserve fairness.

To see how many copies we need in the worst case, note that every process 
is either visited finitely or infinitely often, and from the latter there may be up to $k$ states for which we need two instances. Let's denote $|\visited^{fin}_{B_2...B_n}|$ by $fin$. Then we need at most $fin + 2k + 1$ instances (including one instance for $B_1$). However, if we write $inf$ for $|\visited^{inf}_{B_2...B_n}|$, then we have $fin = |B| - inf$. Then, since we know that $k \leq inf$, we have $fin + 2k + 1 = |B| - inf + 2k + 1 \leq |B| + k + 1$. 
\qed

\subsection{Local Deadlocks, Without Fairness}

We give a bounding lemma for local deadlocks without fairness, using a new construction. 

\begin{lemma}[Bounding Lemma, local deadlocks, disjunctive, non-fair]
\label{lem:bounding-disj-dead-nf}
Let $A,B$ be process templates with $G = G_A \cup G_B$. Let $Q^*_B=\{q \in Q_B \mid |\Enable_q| < |B|\}$ and $m=\max_{q \in Q^*_B}\{|\Enable_q|\}$. Then, for $c =m+|G|+1 \leq n$:
$$(A,B)^{(1,n)} \textrm{ has a local deadlock } \implies (A,B)^{(1,c)} \textrm{ has a local deadlock.}$$
\end{lemma}

\noindent
Note that if a local deadlock is possible in $q$, then $|\Enable_q| < |B|$, i.e., $Q^*_B$ is the set of states in which a local deadlock could occur.

\smartpar{Proof.} Given a locally deadlocked run $x=x_0,x_1...$ of $(A,B)^{(1,n)}$, we construct a locally deadlocked run $y=y_0,y_1...$ of $(A,B)^{1,c}$.\\
\textbf{Construction:}\\
Assume $B_1$ is locally deadlocked in state $q_l$ (other cases are similar):

\begin{enumerate}
\item set $y(A) = x(A)$ and $y(B_1) = x(B_1)$
\item for every $q \in \Enable_{q_l}$, if $q$ appears in the run 
(i.e., $\exists j,m:\: x_m(j)=q$), devote one process $B_{i_q}$ such that $y
(B_{i_q})=x(B_{
\first_q})[1:f_q].(q)^{l_q-f_q}.x(B_{\last_q})[l_q+1:]$
\item for every guard $g \in G$, let $(x(B_i),q_r)$ be the \emph{
representative} for $g$, then devote one process $B_j$ of $(A,B)^{(1,c)}$ 
such that: $y(B_j) = x(B_i)[1:f_{q_r}](q_r)^w$. Note that if $q_r \in 
\Enable_{q_l}$ then we must choose the next \emph{representative} of $g$. If we 
can not find a \emph{representative} that is not in $\Enable_{q_l}$, then we 
simply disregard the guard.
\end{enumerate}
Suppose the deadlock in the original run occured at time $d$, then the 
construction ensures that, at any time $t \geq d$ we have $\neg\exists q_i \in 
\Enable_{q_l}$ and $q_i \in y(t)$. Therefore the local deadlock is preserved 
in the constructed run $y$ at any time greater than $d$. Furthermore, all transitions in $y$ are enabled by a similar argument as in the proof of Lemma~\ref{lem:bounding-disj-nf}. \qed

\subsection{Local Deadlocks, With Fairness}

\begin{lemma}[Bounding Lemma, local deadlocks, disjunctive, fair]
\label{lem:bounding-disj-dead-f}
For process templates $A,B$ with $G = G_A \cup G_B$ and $n \geq |B|+|G|+1$, and strong-fair runs:
$$(A, B)^{(1,n)} \textrm{ has a local deadlock } \implies (A, B)^{(1,|B| + |G| + 1)} \textrm{ has a local deadlock }$$
\end{lemma}

\smartpar{Proof.} We can use the same construction as for Lemma~\ref{lem:bounding-disj-f}, where either process $A$ or process $B_1$ is now the process that is eventually locally deadlocked. The local deadlock is preserved since states that appear finitely often in the original run, also appear also finitely often in the constructed run. Fairness holds by construction.
\qed

\subsection{Global Deadlocks}

For Theorem~\ref{thm:disj}, we defined $\mN = \{q \in Q_B \mid q \in \Enable_q\}$, and $\mN^*$ as the maximal subset (wrt. number of elements) of $\mN$ such that $\forall q_i,q_j \in \mN^*: q_i \notin \Enable_{q_j} \land q_j \notin \Enable_{q_i}$.
To prove the part of the theorem that regards global cutoffs, we need to prove the following lemma.

\begin{lemma}[Bounding Lemma, global deadlocks, disjunctive]
\label{lem:bounding-disj-dead}
For disjunctive systems and $n \geq |B|+|\mN^*|$:
$$(A, B)^{(1,n)} \textrm{ has a global deadlock } \implies (A, B)^{(1,|B|+|\mN^*|)} \textrm{ has a global deadlock }$$
\end{lemma}


\smartpar{Proof.} For a state $q \in Q_A \cup Q_B$, let $\dead^\lor_q = Q_A \cup Q_B \setminus \Enable_q$.
Given a run $x=x_0,x_1...$, a state $q \in Q_B$ is disabled at time $m$ if all of the following hold:
\begin{itemize}
\item[-]  $q \in Set_m(x)$,
\item[-]  $Set_m(x) \setminus \{q\} \subseteq \dead^\lor_{q}$, and
\item[-] if $q \in \Enable_q$ then $\occurs_{x_m}(q) = 1$.
\end{itemize}
A state $q \in Q_A$ is disabled at time $m$ if the first two conditions above hold.
%
Then, a run $x$ is globally deadlocked at time $m$ iff all $q \in Set_m(x)$ are disabled at time $m$.
Note that this holds iff the following two conditions hold:
\begin{itemize}
\item[-] $\forall q_i \neq q_j \in Set(x_m): q_i \in \dead^{\lor}_{q_j}$ and $q_j \in \dead^{\lor}_{q_i}$,
\item[] and
\item[-] $\forall q_i \in \left(Set_m(x) \cap \mathcal{N}\right): \occurs_{x_m}(q_i) = 1$.

\end{itemize}
These conditions determine the configurations of a system $\largesys$ in which a global deadlock is possible. This observation is crucial to obtain smaller cutoffs for global deadlock detection.

The cutoff obtained previously was $c=2|B| - 1$. In the proof of this 
result~\cite{AusserlechnerJK15}, the processes are divided into two sets: 
$\mathcal{C}$ and $\mathcal{B \setminus C}$, where $\mathcal{B}$ is the 
set of all $B$-processes and $\mathcal{C}$ is the set of processes 
deadlocked in a state from $\mN$. In the following, let 
$\visited^{inf}_{\mathcal{B \setminus C}}$ be the set of states 
in which the processes from $\mathcal{B \setminus C}$ are deadlocked, 
and let $\visited^{fin}_{\mathcal{B \setminus C}}$ be the states 
that are only visited on the path to the 
deadlock. Then, a run of $\cutoffsys$ is constructed as follows:
\begin{enumerate}
\item Copy (in addition to process $A$) all local runs of processes in 
$\mathcal{C}$.
\item Flood all deadlocked states of processes $\mathcal{B \setminus C}$
, i.e., that are in $\visited^{inf}_{\mathcal{B \setminus C}}$.
\item All remaining states that appear in the processes $\mathcal{B 
\setminus C}$, i.e., that are in $\visited^{fin}_{\mathcal{B 
\setminus C}}$, are flooded with evacuation into $\visited^{inf}_{
\mathcal{B \setminus C}}$. 
\end{enumerate}
\noindent
Therefore,  $|\mathcal{C}| + |\visited^{fin}_{\mathcal{B \setminus C}}|+ 
|\visited^{inf}_{\mathcal{B \setminus C}}|$ is a cutoff. Since $|\visited^{fin}_{\mathcal{B \setminus C}}| + |\visited^{inf}_{\mathcal{B \setminus C}}| \leq |B|$, also $|\mathcal{C}| + |B|$ is a cutoff. Thus, we can obtain cutoffs smaller than $2|B|-1$ in case $|\mC|$ is smaller than $|B|$. Indeed, we know that $|\mC| \leq |\mN|$, which is in many cases much less than the size of 
$B$. Thus, the cutoff can be reduced to 
$|\mathcal{N}| + |B|$.
If we consider in addition to the properties of single states also the properties of pairs of states, then the cutoff can be minimized further: if two states are not in the $\dead^\lor$ sets of each other, they can never be together part of a global deadlock. Thus, a sufficient size for any subset of $\mN$ that can be in a global deadlock together can be found by computing the maximal subset $\mN^* \subseteq \mN$ such that $\forall q_i,q_j \in \mN^*: q_i \notin \Enable_{q_j} \land q_j \notin \Enable_{q_i}$.\qed

\medskip
\noindent
\paragraph{Remark.} Computing $\mN^*$ exactly amounts to computing the $minimal \; vertex \; cover$ $mvc$ of the undirected graph $G=(V, E)$, where:
\begin{itemize}
\item $V = \mathcal{N}$
\item $E = \{(q_1, q_2) \mid q_1 \not\in \dead^{\lor}_{q_2} \}$
\end{itemize}
The $vertex \; cover$ problem is \emph{NP-Complete}, but it can be safely underapproximated in the following way:
first we sort the states by their number of edges in descending order, then starting from the top, we compute minimum number of states $U$ such that the sum of their edges is greater or equal to $|E|$. The correctness of this method stems from the fact that any set of states with size less than $U$ can never be a $vertex \; cover$.

\section{Appendix: Proofs and Proof Methods for Conjunctive Systems}
\label{sec:proofs-conj}

In this section, we present lemmas and proof methods that allow us to obtain our cutoff results for local and global deadlock detection in conjunctive systems. For \LTLmX properties, we do not give new cutoff results, since the existing ones are already optimal (see Table~\ref{tab:cutoffs-conj}).

\subsection{Definitions}
Given a system $(A, B)^{(1,n)}$, we define the following:
\begin{itemize}
\item We say that $D\subseteq (Q_A \cupdot Q_B)$ is a \emph{deadset} of $q \in (Q_A \cupdot Q_B)$  if $\forall (q,\sigma,g,q') \in \trans: \exists q'' \in D: q'' \notin g$ and  $\forall q'' \in D$ $ \exists (q,\sigma,g,q') \in {\trans}: q'' \not\in g$, and $D$ contains at most one state from $Q_A$. 

\item $dead^\land_q$ is the set of all deadsets of $q$: $dead^\land_q = \{ D \subseteq (Q_A \cupdot Q_B) \mid D \textrm{ is a deadset of } q\}$.

\end{itemize}

\subsection{Global Deadlocks}

Recall that $dead^\land_q = \emptyset$, then we say $q$ is \emph{free}. If a state $q$ does not appear in any $dead^\land_{q'}$, then we say $q$ is \emph{non-blocking}. If a state $q$ does not appear in $dead^\land_q$, then we say $q$ is \emph{not self-blocking}. 

\begin{lemma}[Bounding Lemma, global deadlocks, conjunctive]
\label{lem:bounding-conj-dead}
In a conjunctive system, where 
\begin{itemize}
\item $D_1 \subseteq Q_B$ is the set of free states in $B$, 
\item $D_2 \subseteq Q_B$ is the set of non-blocking states in $B$, and 
\item $D_3 \subseteq Q_B$ is the set of not self-blocking states in $B$.
\end{itemize}
Let $c=2|B|-2|D_1|-2|D_2 \setminus D_1|-| D_3 \setminus (D_1 \cup D_2)|$. Then, for $n \geq c$:
$$(A, B)^{(1,n)} \textrm{ has a global deadlock } \implies \cutoffsys \textrm{ has a global deadlock }$$
\end{lemma}
\smartpar{Proof.}
Given a run $x=x_0,x_1...$, a state $q \in Set_m(x)$ is disabled at time $m$ iff:
\begin{itemize}
\item[-] $\exists D \in dead^{\land}_{q}:$ $D \subseteq Set_m(x)$
\item[-] if $q \in D$  then $occurs_{x_m}(q) \geq 2$.
\end{itemize}
A run $x$ is globally deadlocked at time $m$ iff all $q \in Set_m(x)$ are disabled.

For a deadlocked run $x$ of $\largesys$, let $\visited^{inf} = Set_m(x) \cap Q_B$, i.e., the set of states of $B$ that appear in the deadlock. Au{\ss}erlechner et al.~\cite{AusserlechnerJK15} have shown that then the global deadlock can be replicated in $\cutoffsys$ by copying, for each $q \in \visited^{inf}$, at most two local runs that end in $q$. Since \init is assumed to appear in every guard, the resulting cutoff is $2|B|-2$.

By a similar argument as for \init, we can obtain an even smaller cutoff if any of the other states in process template $B$ satisfy one of the properties defined before this lemma. In particular, \init is an example of a non-blocking state. If there are other non-blocking states in $B$, then the cutoff can be reduced by the same argument as for \init: since such states do not block any transitions, local runs that end in these states can just be removed from the system, and the run will still be deadlocked. Moreover, we can also reduce the cutoff if there are states that are not self-blocking: the reason why we may need $2$ copies of a state $q$ is that the second copy may be needed to block a transition of another process that also is in $q$. However, if $q$ is not self-blocking, then this second copy is not necessary. Finally, if $q$ is free, then $q$ cannot be part of a deadlocked configuration at all, since $q$ always has at least one transition that can be taken. Thus, copied local runs for free states will never be necessary.

Thus, we can reduce the cutoff to $2|B|-2|D_1|-2|D_2 \setminus D_1|-| D_3 \setminus (D_1 \cup D_2)|$. Note that if this results in a cutoff of $0$ or $1$, then we have statically detected that a global deadlock is not possible. \qed
\begin{example}
Consider the process templates in Figure~\ref{fig:dead-example}.

\begin{figure}[h]
\fboxrule=0pt
\fboxsep=0pt
\noindent\fbox{%
\begin{minipage}[t]{0.48\linewidth}
A.\\
  \begin{tikzpicture}[node distance=1.5cm,>=stealth,bend angle=30,auto]
	
  \tikzstyle{state}=[circle,thick,draw=black,minimum size=10mm]

  \begin{scope}
    
    \node [state] (init)  {$in_A$}
    edge [pre, loop right]   node[right=1pt] {\tiny{$\forall \neg 1_B$}} (init)
    edge [pre, loop left]   node[left=1pt] {\tiny{$\forall \neg 2_B$}} (init)
    edge [pre, loop below]   node[below=1pt] {\tiny{$\forall \neg 3_B$}} (init); 
   
  \end{scope}

  \end{tikzpicture}
\end{minipage}
}
\hfill%
\fbox{%
\begin{minipage}[t]{0.48\linewidth}
B.\\
\begin{tikzpicture}[node distance=2.75cm,>=stealth,auto]
  \tikzstyle{state}=[circle,thick,draw=black,minimum size=7mm]

  \begin{scope}
    
    \node [state] (init) {$in_B$};   
    
    \node [state] (b1) [left of=init] {$1_B$}
     edge [pre, bend left = 30]   node[near end,below = 1pt] {}     (init)
     edge [post]   node[below = 0.25pt] {\tiny{$\forall \neg 1_B$}}     (init)
     edge [post, bend right = 30]   node[below = 0.35pt] {\tiny{$\forall \neg 2_B$}}     (init)
     edge [post, bend right = 60]   node[below = 0.5pt] {\tiny{$\forall \neg 3_B$}}     (init);

    \node [state] (b2) [above right of=init]{$2_B$}
     edge [pre, bend left = 25]    node[near end,below=1pt] {}     (init)
     edge [post]   node[above = 0.25pt] {\tiny{$\forall \neg 1_B$}}     (init)
     edge [post, bend right = 30]   node[above = 0.35pt] {\tiny{$\forall \neg 2_B$}}     (init)
     edge [post, bend right = 60]   node[above = 0.5pt] {\tiny{$\forall \neg 3_B$}}     (init);
     
    \node [state] (b3) [below right of=init] {$3_B$}
	 edge [pre, bend left = 30]    node[near end,below=1pt] {}     (init)
     edge [post]   node[below = 0.25pt] {\tiny{$\forall \neg 1_B$}}     (init)
     edge [post, bend right = 30]   node[below = 0.35pt] {\tiny{$\forall \neg 2_B$}}     (init)
     edge [post, bend right = 55]   node[below = 0.5pt] {\tiny{$\forall \neg 3_B$}}     (init);
   
  \end{scope}

  \end{tikzpicture}
\end{minipage}}%
\caption{Example Process templates}
\label{fig:dead-example}
\end{figure}
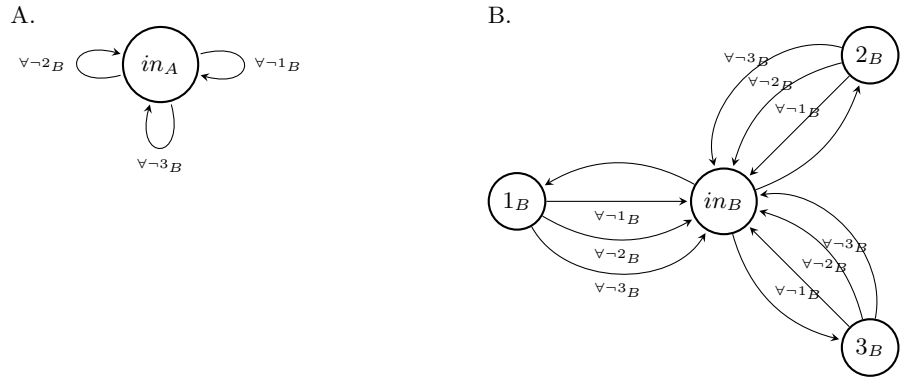
The deadsets of the local states are:\\
$dead_{1_B} =  \{\{1_B, 2_B, 3_B\}\}$\\
$dead_{2_B} = dead_{1_B}$\\
$dead_{3_B} = dead_{1_B}$\\
$dead_{in_A} = dead_{1_B}$\\
$dead_{in_B} = \emptyset$\\

The state $in_B$ can not be part of any global deadlock because its deadset is empty. On the other hand the deadset $D=\{1_B, 2_B, 3_B\}$ can be a part of a global deadlock and it is reachable.
According to the definition of the global deadlock all the states of this set must be duplicated in the run.
\end{example}

\subsection{Local Deadlocks}


Local deadlock detection in conjunctive systems is not an easy task even for the unfair case. The main problem is to find the minimum number of processes needed that can provide an infinite behavior while preserving the deadlock. In some special cases, This number can be found by fetching special \emph{lassos} from the process templates. 

\subsubsection{Definitions}
Given a system $(A, B)^{(1,n)}$ and a run $x=x_1,x_2,\ldots$, we define the following:
\begin{itemize}
\item a sequence of states $q_1 \ldots q_n$ is \emph{connected} if $\forall q_i \in \{q_1, \ldots ,q_n\}: $  $\exists (q_i,\sigma,g,q_{i+1}) \in {\trans}$
\item A \emph{cycle} is a connected sequence of states $q\, q_1 \ldots q_n\, q$ such that $\forall q_i,q_j \in \{q_1, \ldots, q_n\}:$ $q_i \neq q_j$. We denote such a cycle by $C_q$. (By abuse of notation, $C_q$ is also used for the set of states on $C_q$.) We denote the set of guards of the transitions on $C_q$ as $G_{C_q}$

\item A \emph{cycle} $C_q$ is called \emph{free} if $\forall p \in C_q \setminus q$ $\forall g \in G_{C_q}:$ $p \in g$. We denote such a cycle by $C_q^{free}$.
\ms{can a and b share the same cycle?}

\item A \emph{covered alternation} between two states $p$ and $q$ occurs iff $\exists m,m'$ where $m + 1 < m'$, $p \in x_m$, $q \not\in x_m$, $\forall i \in [m + 1,m'[$ $\{p,q\} \subseteq x_i $, $p \not\in x_{m'}$ and $q \in x_{m'}$.
\item A \emph{lasso} $lo$ is a connected sequence of states $q_0 \ldots q_i \ldots q_n$ such that:
\begin{itemize}
\item $q_0$ is an initial state
\item $q_i = q_n$, and $q_i \ldots q_n$ is a cycle.
\end{itemize}
We denote by $G_{lo}$ the set of guards of the transitions on $lo$.
\end{itemize}

\subsubsection{Local Deadlocks in \emph{1-conjunctive} Systems}

\begin{lemma}[Bounding Lemma, local deadlocks, 1-conjunctive, non-fair]
\label{lem:bounding-1-conj-dead-nf}
For a \emph{1-conjunctive} system $(A, B)^{(1,n)}$ and $n \geq |G_B| + 2$:\\
 $$(A,B)^{(1,n)} \textrm{ has a local deadlock } \implies (A,B)^{(1,|G_B| + 2)} \textrm{ has a local deadlock}$$ 

\end{lemma}
\smartpar{Proof.}
This result follows from Au{\ss}erlechner et al.~\cite[Lemma 12]{AusserlechnerJK15}. The proof construction in a 
nutshell was that if in a run of $\largesys$, process $B_1$ is locally deadlocked in some state $q_l$ at 
time $d$, then we construct a run of $\cutoffsys$ by computing $q_l$'s deadset and for each state $q \in Q_B$ in the deadset we copy one local run until it visits $q$, and then we let it stay in $q$ forever. In addition, we copy the local runs of $B_1$ and some process that 
moves infinitely often. Since our system is \emph{1-conjunctive}, the size of any \emph{deadset} 
is always less or equal to $|G_B|$. 

\begin{lemma}[Bounding Lemma, local deadlocks, 1-conjunctive, fair]
\label{lem:bounding-1-conj-dead-f}
For a \emph{1-conjunctive} system $(A, B)^{(1,n)}$ and $n \geq 2|G_B| + 1$ and strong-fair runs:

 $$(A,B)^{(1,n)} \textrm{ has a local deadlock } \implies (A,B)^{(1,2|G_B| + 1)} \textrm{ has a 
local deadlock}$$

\end{lemma}
\smartpar{Proof.}
Similar to what we have described above, we get this result by inspection of the proof of Au{\ss}erlechner et al.~\cite[Lemma 16]{AusserlechnerJK15}. The original construction includes $2$ local runs for every state in the deadset, and one additional state that is locally deadlocked. Since the size of the deadset is bounded by $|G_B|$, we get that $2|G_B|+1$ processes are sufficient to replicate the local deadlock.
\qed

\subsubsection{Local Deadlocks: Beyond \emph{1-conjunctive} Systems}

In this sections we will show how to obtain cutoffs for conjunctive systems that are not $1$-conjunctive. First, we will consider a number of cases that can be reduced the $1$-conjunctive case, and therefore have the same cutoff. Then, we will consider a case that cannot be reduced to the $1$-conjunctive case, and show that it already requires a significantly larger cutoff. Example~\ref{ex:quadratic-cutoff} shows that the cutoff for local deadlock detection in general conjunctive systems is at least quadratic in the number of states, and can grow independently of the number of guards. Since a general cutoff results are very hard to obtain, and would not be very useful because of their size, we restrict ourselves to these partial results.

Below, for simplicity we explain one case in detail: a system $(A, B)^{(1,n)}$ where a single guard, say $(g_{q_l}^2 = \neg a \land \neg b)$, is $2$-conjunctive, and all other guards are \emph{1-conjunctive}. We further assume that $g_{q_l}^2$ only appears in transitions from $q_l$ to some other state. 

We then explain how this case can be generalized.

\paragraph*{Systems with Alternation-free Local Deadlocks}

If any of the following holds, then for non-fair runs we can reduce the problem to the $1$-conjunctive case:

\begin{enumerate}
\item If the deadlock is not possible on $q_l$, either because $q_l$ is not reachable, or because $q_l$ is free. Since we assumed that all other processes have only $1$-conjunctive guards, the problem reduces to the $1$-conjunctive case, and the same cutoff applies. This also holds for the fair case.
\item If there exists a \emph{lasso} $lo_1$ such that $\forall (q_l,\sigma,g,q') \in {\trans}$, $\forall g_{lo_1} \in G_{lo_1}$ we have $g_{lo_1} \neq \neg q_l$ and $g_{lo_1} \cap g = g_{lo_1}$ then the $1$-conjunctive cutoff applies.

The idea of this restriction is that we need one process that can move infinitely often, after the deadlocked process enters $q_l$ and we have other processes in all the states that disable $q_l$. Since one representative per guard is enough for this, we need at most $|G_B|$ processes to disable $q_l$. The two additional processes are the deadlocked one and the one that moves through the lasso. This process waits in $\init$ until all other processes have reached their destination. Then, by construction, it can take transitions along the lasso until infinity. Since this construction is inherently not fair, it does not give a cutoff for the fair case.

\item The requirement above can be relaxed, in that not a single lasso must be free of both $\neg a$ and $\neg b$, but it is sufficient if two separate lassos exist, one that is free of $\neg a$, and one that is free of $\neg b$:

If there exist two \emph{lassos} $lo_1$ and $lo_2$ such that $\forall (q_l,\sigma,g,q') \in {\trans}$, $\forall g_{lo_1} \in G_{lo_1}$ we have $g_{lo_1} \neq \neg q_l$ and $a \in (g_{lo_1} \cap g)$ and $\forall (q_l,\sigma,g,q') \in {\trans}$, $\forall g_{lo_2} \in G_{lo_2}$ we have $g_{lo_2} \neq \neg q_l$ and $b \in (g_{lo_2} \cap g)$, then the $1$-conjunctive cutoff applies.
 
The cutoff does not increase compared to the previous case, since the construction will only use one of the lassos, depending on whether $a$ or $b$ are present in the local deadlock state of the other processes. Again, the construction is inherently not fair.
\item If for all cycles $C_a$ that traverse $a$ we have $G_{C_a} \cap (G_{q_l} \cup \neg q_l) \neq \emptyset$, or for all cycles $C_b$ that traverse $b$ we have $G_{C_b} \cap (G_{q_l} \cup \neg q_l) \neq \emptyset$, or if we have $G_{q_l} \cap \{\neg a, \neg b\} \neq \emptyset$, then the cutoff for $1$-conjunctive systems applies both in the non-fair and the fair case.

The idea is that under each of this assumptions, an infinite alternation between $\{a, \neg b\} \in x_i$ and $\{\neg a, b\} \in x_i$ is not possible. Then we simply copy one process for every $1$-conjunctive guard of $q_l$, and one process for either $a$ or $b$, as well as one more process that moves infinitely often in the original run. 

For the fair case, we need up to $2$ processes to ensure that every process that is enabled can also move eventually, similar to the $1$-conjunctive fair case.
\end{enumerate}

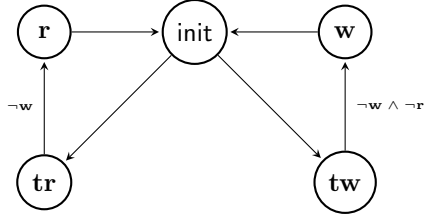
\begin{example}~\\
\begin{tikzpicture}[node distance=2cm,>=stealth,auto]
  \tikzstyle{state}=[circle,thick,draw=black,minimum size=7mm]

  \begin{scope}
    
    \node [state] (init) {$\init$};   
    
    \node [state] (r) [left of=init] {$\mathbf{r}$}     
     edge [post] (init);
     
    \node [state] (tr) [below of=r] {$\mathbf{tr}$}     
     edge [pre] (init)
     edge [post] node[left = 0.5pt] {\tiny{$\mathbf{\neg w}$}} (r);
     
    \node [state] (w) [right of=init] {$\mathbf{w}$}     
     edge [post] (init);    
     
    \node [state] (tw) [below of=w] {$\mathbf{tw}$}     
     edge [pre] (init)
     edge [post] node[right =0.5pt] {\tiny{$\mathbf{\neg w \land \neg r}$}}(w);

  \end{scope}
  \end{tikzpicture}
  \begin{itemize}
  \item if the local deadlock is in a node that has no \emph{2-conjunctive} guard, then the problem is reduced to \emph{1-conjunctive} system.
  \item if the local deadlock is in $tw$, and as all the cycles that contain $w$ contain also $tw$ then the \emph{covered alternation} is not possible. But as there is a lasso $lo = [init,tr,r,init]$ that is free of the guard $\neg r$, then the \emph{1-conjunctive} cutoff for local deadlock detection can be used.
  \end{itemize}
\end{example}

The special cases above can be generalized in the following way, which in many cases results in strong restrictions on the process template:
\begin{enumerate}
\item If the deadlock is not possible in any state that has guards that are not $1$-conjunctive (either because they are not reachable, or because they are free), then the problem reduces to the $1$-conjunctive case, and the same cutoff applies.
\item If for every state $q_l$ with a set of transitions with not $1$-conjunctive guards $G_{q_l} = \{g_1,\ldots,g_n\}$, there exists a \emph{lasso} $lo_1$ such that $\forall (q_l,\sigma,g,q') \in {\trans}$, $\forall g_{lo_1} \in G_{lo_1}$ we have $g_{lo_1} \neq \neg q_l$ and $g_{lo_1} \cap g_i = g_{lo_1}$ for all $g_i \in G_{q_l}$, then the $1$-conjunctive cutoff applies.

The idea of this restriction is a straightforward generalization of what is described above.
\item As above, we can have several lassos instead of a single one: if for every state with a set of not $1$-conjunctive guards $G_{q_l} = \{g_1,\ldots,g_n\}$, and for every state $q \in g_i$, there exists a \emph{lasso} that is free of $\neg q_l$ and $\neg q$, then the $1$-conjunctive cutoff applies.

\item Similar to what we had for the lassos, for every state $q_l$ with a set of transitions with not $1$-conjunctive guards $G_{q_l} = \{g_1,\ldots,g_n\}$, for every $g_i = \neg q_1 \land \ldots \land \neg q_k$ there must exist $k-1$ cycles that are not traversable during the local deadlock. In this case, we know exactly which states can appear infinitely often during a local deadlock, and the cutoff for $1$-conjunctive systems applies in both the non-fair and the fair case.
\end{enumerate}

\paragraph{Systems without Alternation-free Local Deadlocks}

In this section we will assume that special cases do not hold, and we have to consider the case that, for a $2$-conjunctive guard $g = \neg a \land \neg b$, alternating infinitely often between $a$ and $b$ is necessary to obtain a locally deadlocked run.

We need the following additional definitions:
\begin{itemize}
\item A \emph{segment} $Sg_{a-b}$ is a connected sequence of states $q_i \ldots q_j$ where:
\begin{itemize}
	\item $q_i$ has an incoming transition with guard $\neg a$
	\item $q_j$ has an outgoing transition with guard $\neg b$
	\item $\forall q_m \in Sg_{a-b}$ $\exists (q_m,\sigma, g, q_{m+1}) \in \delta:$ if $q_{m+1} \in Sg_{a-b}$ then $ b \in g$
\end{itemize}
\item For a cycle $C_q$, we denote by $|Sg_{a-b}|_{C_q}$ the total number of segments $Sg_{a-b}$ on $C_q$
\item A \emph{segment transition} on some cycle $C_x$ is a path $(s_1,e_1,p)(s_2,e_2,p)\ldots(s_n,e_n,p)$ such that $s_1(p) \in Sg_{a-b}$ and $s_n(p) \in Sg_{b-a}$ and $\forall i$ $s_i(p) \in C_x$ and $\exists p' \neq p:$ $s_1(p') = a$ and $b \not\in s_1$. 
\end{itemize}

For systems with a single $2$-conjunctive guard that need to alternate between $a$ and $b$ to obtain a local deadlock, we state the following.
\begin{lemma}
\label{lem:1-2-conj}
Given a single \emph{2-conjunctive} system $(A, B)^{(1,n)}$ deadlocked locally in state $q_l$, $(g_{q_l}^2 = \neg a \land \neg b)$, with unique cycles $C_a$ and $C_b$ where these cycles are free and $G_{(C_a \cup C_b)} \cap (G_{q_l} \cup \neg q_l) = \emptyset$. Let 
$$n_a = max(|Sg_{a-b}|_{C_a}  ,|Sg_{b-a}|_{C_a})$$
$$n_b = max(|Sg_{a-b}|_{C_b}  , |Sg_{b-a}|_{C_b}).$$
Then: 
 $$(A,B)^{(1,n)} \textrm{ has a local deadlock in $q_l$} \implies (A,B)^{(1,|G_B| + n_a + n_b + 5)} \textrm{ has a local deadlock in $q_l$}.$$ 
\end{lemma}

To prove the lemma, we will use the following observation on transitions between segments on free cycles.

\begin{obs}\label{thr:alter}
Given a single \emph{2-conjunctive} system $(A, B)^{(1,n)}$ deadlocked locally in state $q_l$, $(g_{q_l}^2 = \neg a \land \neg b)$, if there exist two cycles $C_a^{free}$ and $C_b^{free}$ where $\forall g \in G_{C_a^{free}}:$ $C_b^{free} \setminus b \subseteq g$ and $\forall g \in G_{C_b^{free}}:$ $C_a^{free} \setminus a\subseteq g$ then at any moment $m$, if $Set(x_m) \subseteq (C_a^{free} \cup C_b^{free})$ then:

$$\textrm{ if } a \in x_m \textrm{ and  } b \not \in xm \textrm{ then: }$$
$$\exists \textrm{ segment transition } Sg_{a-b} \textrm{ to } Sg_{b-a}$$
$$\neg\exists \textrm{ segment transition } Sg_{b-a} \textrm{ to } Sg_{a-b}$$

$$\textrm{ if } b \in x_m \textrm{ and  } a \not \in xm \textrm{ then :}$$
$$\exists \textrm{ segment transition } Sg_{b-a} \textrm{ to } Sg_{a-b}$$
$$\neg\exists \textrm{ segment transition } Sg_{a-b} \textrm{ to } Sg_{b-a}$$

\end{obs}

\smartpar{Proof of Lemma~\ref{lem:1-2-conj}.} First we need to prove that if the number of processes on $C_a$ is less than $n_a +1$, then the deadlock cannot be preserved. Suppose we have $n_a$ processes on $C_a$ at some time $m$, we distinguish three cases:
\begin{enumerate}
\item All processes are in $Sg_{b-a}$ and $a \in Sg_{a-b}$: In this case $b \in x_m$. According to Observation~\ref{thr:alter}, all processes can make a \emph{segment transition}, then at some time $m'$, assuming all processes move whenever possible, all processes are in $Sg_{b-a}$ and in particular a process must be in $a$. Now after another \emph{covered alternation}, all processes can make a \emph{segment transition} except the one in $a$, then the number of processes in $Sg_{b-a} = |Sg_{b-a}|_{C_a} - 1$, then at some point in time $> m'$, by pigeonhole principle, neither $a$ nor $b$ will be covered and thus the deadlock can not be preserved.
\item All processes are in $Sg_{a-b}$ and $a \in Sg_{b-a}$: similar argument to the above.
\item Processes are scatered between $Sg_{b-a}$ and $a \in Sg_{a-b}$: If this was the case and as we only have $n_a$ process, then we will have at least two empty consecutive segments $Sg_{b-a}$ and $Sg_{a-b}$ then at some time in the future a \emph{covered alternation} is not possible.
\end{enumerate}
We can deduce from the above that at least $n_a + 1$ processes can reach $C_a$ and at least $n_b + 1$ processes can reach $C_b$. Note that we might need one additional process for $C_a$ cycle if $\exists q_1,q_2$ in $a$'s \emph{segment} where these two states have outgoing transitions on the cycle with guard $\neg b$ and one of them appears before $a$ and the other after it(same applies for $C_b$). In the following we will denote by $k_a$ either $n_a + 1$ or $n_a + 2$ and by $k_b$ either $n_b + 1$ or $n_b + 2$, depending whether the special case applies or not.\\
\smartpar{Construction.}
Given a run $x=x_1,x_2,\dots$, let the process $B_1$ be the deadlocked process in state $q_l$, we construct the run $y=y_1,y_2,\ldots$ as follows:
\begin{itemize}
\item $y(B_1) = x(B_1)$
\item let $D \in dead^\land_q$ then $\forall q \in D \setminus \{a,b\}:$ $y(B_{i_q}) = x(B_{first_q})[1,f_q](q)^{\omega}$
\item $\exists m_1,\ldots,m_{k_a}$ where $x_{m_i}(B_{m_i}) = q:$ $q \in C_a$ then $y(B_j) = x(B_{m_i})[1:m_i]$
\item $\exists t_1,\ldots,t_{k_b}$ where $x_{t_i}(B_{t_i}) = q:$ $q \in C_b$ then $y(B_u) = x(B_{t_i})[1:t_i]$
\end{itemize}
\smartpar{Starting Positions}
\begin{itemize}
\item let all processes move outside $a$ or $b$
\item for each \emph{segment} $Sg_{b-a}$ in $C_a$ or $C_b$ let one process reachs it
\item let remaining processes in the closest position to $a$ or $b$
\end{itemize}

\smartpar{Infinite Behavior Loop.}
In the following loop, we require that no process leaves the cycle that was assigned for it in the start position. 
\begin{enumerate}

\item let a single process moves into $b$
\item leave $a$
\item let all processes take all possible transitions except those that enters $a$ or $b$
\item let a single process moves into $a$
\item leave $b$
\item let all processes take all possible transitions except those that enters $a$ or $b$
\item go to 1
\end{enumerate}

Starting positions are valid as we assumed that the cycles are free and their guards are independents of both cycles states. The infinite behavior loop chosen ensures continues \emph{covered alternation} between $a$ and $b$, this is due to the fact that the loop has the following two invariants:
\begin{enumerate}
\item At anytime $m$ after starting the loop, there is always a process in $a$, or a process with enabled transitions to reach $a$ (while b is occupied).
\item At anytime $m$ after starting the loop, there is always a process in $b$, or a process with enabled transitions to reach $b$ (while a is occupied).
\end{enumerate}\qed

\section{Appendix: Proofs and Proof Methods for Extensions}

\begin{lemma}[Bounding Lemma for Disjunctive \textsc{Target}]
For disjunctive systems and process templates $A,B$ with $q \in Q_B$:

$$(A, B)^{(1,n)} \models \textsc{Target}(q) \implies (A, B)^{(1,|B|)} \models \textsc{Target}(q)$$

\end{lemma}

\paragraph{Proof.}
Given a run $x$ of $\largesys$ where eventually all $B$-processes are in $q$ at the same time $m$, let $D\subseteq Q_B$ be the set of all states of $B$ that appears in $x$ up to time $m$. To construct a run $y$ that satisfies $\textsc{Target}(q)$ in $(A, B)^{(1,|B|)}$, we flood all states in $D$, and evacuate them to $q$ at the time they occur for the last time before moment $m$. Since neither flooding of a state, nor evacuation from a state can depend on another process in the same state, $|B|$ processes are sufficient, at most one per state.
\qed

\begin{lemma}[Bounding Lemma for Disjunctive \textsc{Repeat-Target}]
For disjunctive systems and process templates $A,B$ with $q \in Q_B$:

$$(A, B)^{(1,n)} \models \textsc{Repeat-Target}(q) \implies (A, B)^{(1,|B|)} \models \textsc{Repeat-Target}(q)$$

This result holds with or without restriction to fair runs.
\end{lemma}

\paragraph{Proof.}
To construct a run $y$ of $(A, B)^{(1,|B|)}$, we essentially use the construction from above twice. The construction is the same up to moment $m$. Then, in the original run there must be a time $m'$ such that all processes are again in $q$. Let $D'$ be the set of all states that appear between $m$ and $m'$ in $x$, and use the same construction as above to extend the run $y$ until all processes visit $q$ again. This construction can then be repeated to obtain an infinite run that satisfies $\textsc{Repeat-Target}(q)$. To obtain a fair run, we may have to consider not a simple loop from $\forall i. q_i$ to $\forall i.q_i$, but we have to find a loop such that every process moves at least once. If the original run was fair, such a loop must exist. The cutoff remains the same.
\qed

\end{document}